\documentclass[journal,twoside]{asmb}

\begin{document}
\title{Secrecy Outage Analysis of Energy Harvesting Relay-based Mixed UOWC-RF Network with Multiple Eavesdroppers}

\author[1]{Moloy Kumar Ghosh}
\author[2]{Milton Kumar Kundu}
\author[3]{Md Ibrahim}
\author[4]{A. S. M. Badrudduza}
\author[5]{Md. Shamim Anower}
\author[6]{Imran Shafique Ansari}
\author[7]{Ali A. Shaikhi}
\author[8]{Mohammed A. Mohandes}

\affil[1,2]{Department of Electrical \& Computer Engineering, Rajshahi University of Engineering \& Technology (RUET), Rajshahi-6204, Bangladesh}
\affil[3]{Institute of Information \& Communication Technology (IICT), RUET}
\affil[4]{Department of Electronics \& Telecommunication Engineering, RUET}
\affil[5]{Department of Electrical \& Electronic Engineering, RUET}
\affil[6]{James Watt School of Engineering, University of Glasgow, Glasgow G12 8QQ, United Kingdom}
\affil[7,8]{Electrical Engineering Department, King Fahd University of Petroleum and Minerals, Dhahran 31261, Saudi Arabia}

\twocolumn[
\begin{@twocolumnfalse}
\maketitle
\begin{abstract}
\section*{Abstract}

This work deals with the physical layer security performance of a dual-hop underwater optical communication (UOWC)-radio frequency (RF) network under the intruding attempts of multiple eavesdroppers via RF links. The intermediate decode and forward relay node between the underwater source and the destination transforms the optical signal into electrical form and re-transmits it to the destination node with the help of harvested energy by the relay from an integrated power beacon within the system. The source-to-relay link (UOWC) follows a mixture exponential generalized Gamma turbulence with pointing error impairments whereas all the remaining links (RF) undergo $\kappa-\mu$ shadowed fading. With regards to the types of intruders, herein two scenarios are considered, i.e., colluding (\textit{Scenario-I}) and non-colluding (\textit{Scenario-II}) eavesdroppers and the analytical expressions of secure outage probability, probability of strictly positive secrecy capacity, and effective secrecy throughput are derived in closed form for each scenario. Furthermore, the impacts of UOWC and RF channel parameters as well as detection techniques on secrecy capacity are demonstrated, and following this a comparison between the two considered scenarios is demonstrated that reveals the collusion between the eavesdroppers imposes the most harmful threat on secrecy throughput but a better secrecy level can be attained adopting diversity at the destination and power beacon nodes along with heterodyne detection rather than intensity modulation and direct detection technique. Finally, all the derived expressions are corroborated via Monte Carlo simulations.
\end{abstract}

\begin{IEEEkeywords}
\section*{Keywords} 

Effective secrecy throughput, underwater optical communication, secure outage probability, energy harvesting, colluding, and non-colluding eavesdroppers.

\end{IEEEkeywords}
\end{@twocolumnfalse}
]

\section{Introduction}
Communication technology is expanding at a rapid rate along with the diverse connectivity needs of individuals. While the world is implementing the fifth generation ($5$G) of cellular networks, researchers all around the globe have already begun contemplating and subsequently performing research for the $6$G communication networks. According to the researchers, one of the most important features of $6$G communication system will be the spectral and energy efficiencies \cite{dang2020should}. Energy harvesting can play a vital role in achieving the energy efficiency feature in a wireless communication network applicable to $5$G \cite{DBLP:journals/corr/abs-1910-00785} and $6$G as well as unravel a variety of problems that are impossible to solve by the conventional battery-powered communication operations such as untethered mobility, monitoring in rural areas, and developed medical applications \cite{7010878}.

\subsection{Background}
A communication node can harvest energy using radio frequency (RF) propagation. The harvested power changes significantly with the changes in the number of RF sources and channel condition \cite{7374745}. Energy harvesting schemes can be achieved by considering various possibilities (e.g., single/two-hop model, finite/infinite energy capacity, perfect/imperfect channel state information, etc.) and employing the optimal policy (e.g., offline or online optimal policy) \cite{7120021}. Another scheme named energy cooperation save-then-transmit was propounded in \cite{7044596} where the maximum throughput and outage probability (OP) was derived in the closed-form assuming additive white Gaussian noise channel with deterministic energy arrival rate and Rayleigh block fading channel with stochastic energy arrival rate. A general approximation framework was introduced in \cite{7081079} for real-life energy harvesting setups for both single and multiple users to provide an effective solution to the throughput and outage problems. A piece-wise linear approximation model was proposed in \cite{8315145} considering practical harvesting scenarios such as limited harvesting efficiency and sensitivity. This model is proclaimed to match the actual condition whereas the infinite sensitivity model (both linear and non-linear) deviates from reality. The $2$D and $3$D position of the energy harvesters plays a significant role in harvesting energy to different nodes of a wireless network. If we presume a system with concurrent energy transfer, the received and interference power follows the log-normal distribution while the harvested voltage exhibits the Rayleigh distribution \cite{7127689}.

Energy harvesting can also be boosted by using multiple antennas at the transmitter due to its non-linear characteristics, even if the channel state information (CSI) is unavailable at the transmitter \cite{8470248}. Energy harvesting technology is frequently considered in mixed communication networks such as RF/free-space optical (FSO) networks, RF/underwater optical wireless communication (UOWC) networks, etc. because it requires a large amount of energy for converting the signal from one form to another. In \cite{ZHANG2020126309}, the authors analyzed the OP and diversity order of a FSO/RF network considering the physical limitations like pointing error, atmospheric turbulence, and saturation threshold of the energy harvester, whereas in \cite{ODEYEMI2020125219}, the authors considered a RF/FSO network with a multi-antenna source that harvests energy from a relay and determined the OP of the system. The RF/UOWC is another popular model to explore underwater activities such as ocean surveillance and exploration, climate monitoring, etc. \cite{RAMAVATH2020125774}. The researchers in \cite{8746368, 9217969, 9508176, 9145142} considered a RF/UOWC network to evaluate the OP and average bit error rate (ABER) assuming different combinations of RF and UOWC channels. The application of unmanned aerial vehicle (UAV) as a source was considered in \cite{9563233} where the authors analyzed similar performance parameters (OP and ABER) and also deduced the optimal altitude of the UAV for performance maximization. In another research, the authors assumed a RIS-assisted RF/UOWC network and derived the OP, ABER, and average channel capacity of the system \cite{9352962}.

Another important feature of the $6$G wireless communication model is data security and privacy \cite{9231044}. Since the beginning of the wireless network, this issue has been thoroughly analyzed for RF communication systems. With the rise of mixed networks and advanced eavesdropping technologies, the physical layer security of these types of communication schemes is threatened and needs to be evaluated extensively. In \cite{islam2020secrecy}, authors analyzed the secrecy of a RF/FSO network considering a single eavesdropper trying to overhear information from the RF link. The researchers in \cite{8864007} analyzed energy harvested RF/FSO network assuming the relaying protocol to be decode-and-forward (DF) with the simultaneous wireless information and power transfer (SWIPT) technology and revealed that security of the proposed network can be enhanced by lowering the power splitting fraction parameter. The effect of atmospheric turbulence and pointing error on the security of the RF/FSO model was analyzed in \cite{sarker2021intercept, juel2021secrecy}. A multi-relay network was considered in \cite{8226772} with imperfect channel state information (CSI) and non-linear energy harvesters where the authors derived secure outage probability (SOP) to evaluate the system performance. The researchers concluded that further improvement in SOP is impossible if the saturation threshold is higher than a certain value. However, the secrecy performance of a FSO/RF system was investigated in \cite{10.1117/1.OE.60.6.066102} considering an energy harvesting-dependent relay scheme. The authors of \cite{8742355} considered a mixed RF/UOWC model with a single antenna source and receiver in the presence of a relay equipped with multiple antennas and derived the SOP performance of the system. In another work, the authors analyzed three main secrecy parameters i.e., SOP, average secrecy capacity (ASC), and strictly positive secrecy capacity (SPSC) assuming an RF/UOWC model where the underwater signal undergoes frequent underwater turbulence (UWT) due to the temperature gradient, bubble level, and salinity gradient of the ocean \cite{9330523, 9622195}.

The security of the wireless communication model also depends on how the eavesdroppers behave. If there are multiple eavesdroppers in the system, they can work together to decode a message coming from the source. This scenario is known as colluding eavesdropping. On the other hand, eavesdroppers can try to decode confidential information independently without help from other eavesdroppers. This is known as a non-colluding eavesdropping scenario. It is evident that the former scenario is more threatening for wireless communication than the latter \cite{7350251}. Keeping this in mind, many researchers have started to analyze the colluding eavesdropping scenario and how it can jeopardize our secure communication network. The authors in \cite{5206050, 8327926} analyzed the SOP of a wireless network considering colluding eavesdroppers were present in the system while in \cite{9591254, 8076513} the researchers analyzed the same parameter assuming the presence of non-colluding eavesdroppers. To compare the detrimental effect caused by colluding and non-colluding eavesdroppers on the secure wireless network, the authors of \cite{9260703, fi13080205, odeyemi2018physical} considered both eavesdropping scenarios and concluded that the increasing number of eavesdroppers degrades the security of the system whereas colluding eavesdroppers have the most harmful effect.


\subsection{Motivation and Contributions}
Recently, researchers have made some excellent advancements in mixed communication systems. Despite their efforts, a few important considerations such as energy harvesting at the relay node along with the presence of multiple eavesdroppers (both colluding and non-colluding) in the system are not available in any literature and thus are considered an open problem. So, in this work, an energy harvesting relay-based mixed UOWC-RF network with multiple eavesdroppers has analyzed wherein the UOWC channel link follows mixture exponential generalized Gamma (mEGG) fading model whereas the RF link undergoes $\kappa-\mu$ shadowed fading channel. The proposed mEGG model is mostly favored by the researchers due to its capability to mathematically represent all the physical constraints such as air bubbles, UWT, water salinity, and temperature gradient \cite{zedini2019unified, 9917328} along with being a generalized model that can be reduced to EGG model as one of its special cases \cite{illi2019physical}. On the other hand, the $\kappa-\mu$ shadowed fading model is another generalized model that represents a number of practical fading channels \cite[Table 1]{paris2013statistical}. The goal of this research is to analyze the secrecy of such a system considering multiple eavesdroppers present in the system and trying to decode transmitted information individually or as a group as well as measure the impact of energy harvesting on data security. Although a few research has considered colluding \& non-colluding eavesdropping \cite{odeyemi2018physical, shakir2020physical, xia2019secure} and energy harvesting in mixed wireless optical-RF channels \cite{chen2019novel, illi2019physical, saber2019secure}, no research is available considering both of these scenarios. Thus, this research represents a novel system model and provides some unique results. The key contributions of this work are:

\begin{enumerate}

\item At first, we derive the cumulative density function (CDF) of the dual-hop SNR considering an energy-harvested relay for the UOWC-RF network for both source-to-relay as well as eavesdroppers. Note that, unlike \cite{illi2019physical, saber2019secure}, which only considered a single eavesdropper, multiple eavesdropping scenarios (both colluding and non-colluding) is analyzed here. As our model represents a novel structure, the derived CDFs are also novel.

\item The secrecy performance of the proposed network is demonstrated with respect to the secrecy outage probability (SOP), strictly positive secrecy capacity (SPSC), and effective secrecy throughput (EST) expressions for both eavesdropping conditions in closed-form and further verified via Monte-Carlo (MC) simulations. To the best of the authors' knowledge based on the open literature, these expressions are novel and generalized and can be utilized to unify versatile classical existing models as given in \cite{illi2019physical} and \cite[Table 1]{paris2013statistical}.

\item Capitalizing on the derived expressions, noticeable impacts of air bubbles and temperature gradients-based UWTs for both salty and fresh waters along with the energy harvesting parameters are demonstrated. Finally, the effects of two types of detection techniques i.e. intensity modulation/direct detection (IM/DD) and heterodyne detection (HD) techniques are also analyzed.

\end{enumerate}

\subsection{Paper Organization}
The rest of the paper is organized as follows. Section II describes the system model in detail while Section III portrays the channel models. Expression of the performance parameters (SOP, SPSC, and EST) are derived in Section-IV followed by numerical results discussed in Section-V. Finally, concluding remarks on the work are provided in Section-VI.

\section{System Model}
\begin{figure*}[t!]
\vspace{0mm}
    \centerline{\includegraphics[width=0.70\textwidth]{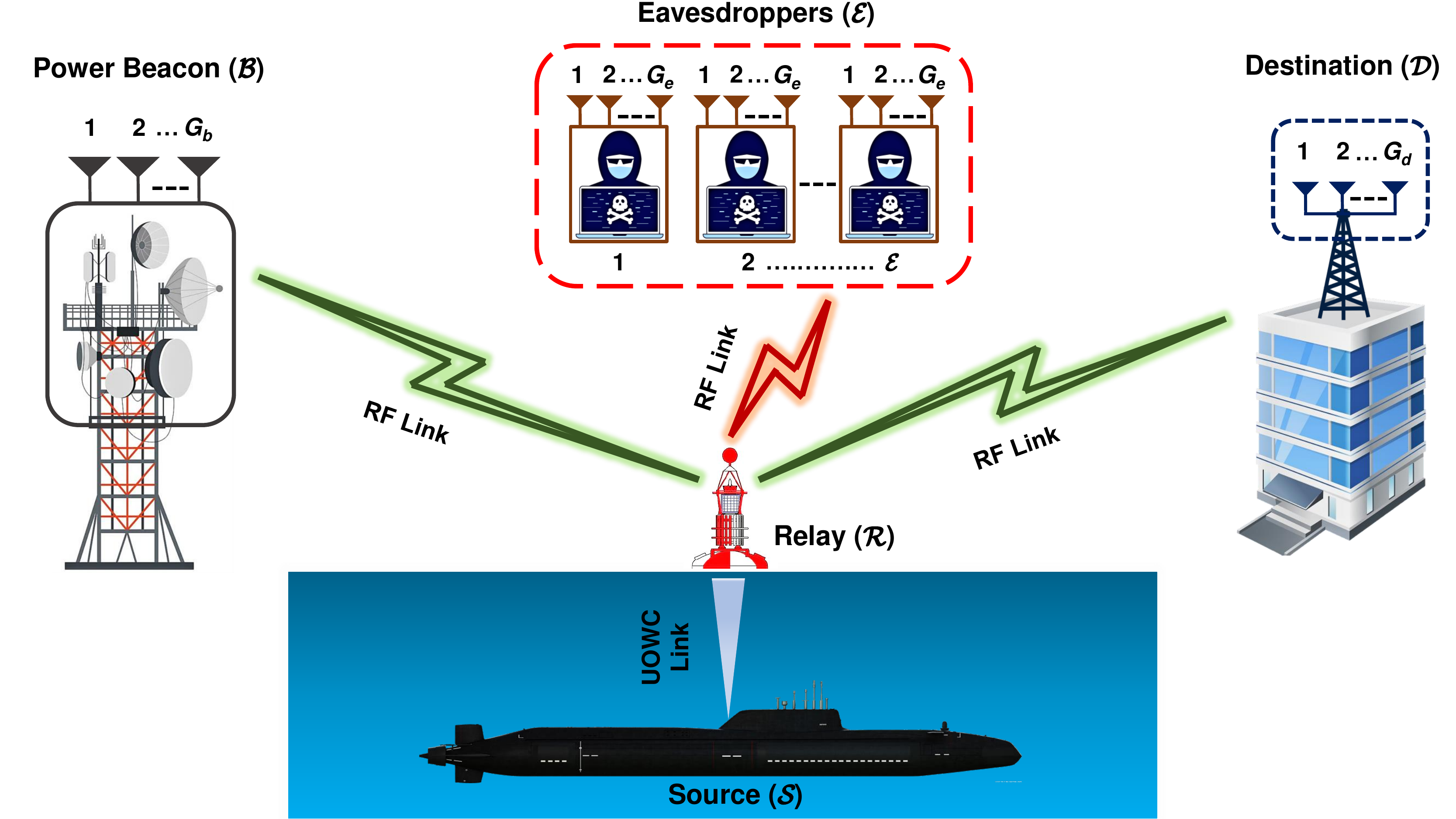}}
    \vspace{0mm}
    \caption{Proposed System Model for Energy Harvesting Relay-based Mixed UOWC-RF Network with Multiple Eavesdroppers.}
    \label{Fig:17}
    \vspace{0mm}
\end{figure*}

In Figure \ref{Fig:17}, an energy-harvested relaying based mixed UOWC-RF system is illustrated that consists of a single aperture source node $\mathcal{S}$ (e.g., submarine, ship), an energy-harvested relay node $\mathcal{R}$ (e.g., tsunami or floating buoy) with a single receive aperture and a single transmit antenna, a destination node $\mathcal{D}$ having $G_d$ antennas, a power beacon $\mathcal{B}$ having $G_b$ antennas, and $\mathcal{E}$ eavesdropper nodes each having $G_e$ antennas.
Herein, $\mathcal{S}$ connected to an underwater optics-based network transmits sensitive information to $\mathcal{D}$ in the RF network while $\mathcal{E}$ tries to steal the sensitive information utilizing the illegitimate RF link. The UOWC link experiences mEGG turbulence with pointing error impairments while all the RF links undergo $\kappa-\mu$ shadowed fading channel. We take into consideration two scenarios that correspond to the two types of eavesdroppers i.e. colluding and non-colluding. In \textit{scenario-I}, the colluding eavesdroppers combine and share their receptions utilizing the maximum ratio combining (MRC) technique in order to decode the private data wiretapped from $\mathcal{R}$. With regard to \textit{scenario-II}, each non-colluding eavesdropper individually examines the wiretapped private data from $\mathcal{R}$. Exploiting the broadcast nature of RF signals, any prototype in the RF system can serve as a power beacon in our proposed framework. Hence, the integration of a power beacon is possible without any additional cost. Since an optical network is unable to directly connect with an RF network, an energy-harvested relay is needed to act as an intermediary to facilitate this communication. It is noteworthy that the decode-and-forward (DF) protocol is utilized to process the received signals at $\mathcal{R}$ since it has the advantage of reducing the first hop's channel impacts on the received signals on the second hop. It is noted that the proposed framework can be utilized in real-world scenarios such as monitoring biological and ecological processes in ocean environments, investigation of climate change, unmanned underwater vehicles utilized to control and maintain oil production facilities, etc.

The entire communication process, $T$, can be divided into two time frames. In the first-time frame, $\mathcal{S}$ transmits the confidential information to $\mathcal{R}$ through the legitimate UOWC link. At $\mathcal{R}$, the instantaneous electrical signal-to-noise ratio (SNR) is expressed as
\begin{align}
   \label{eqn60}
   \gamma_{sr}=\frac{\eta_{sr} ^2 \daleth_{sr} ^2 I_{sr}^2}{\vartheta _{sr}^2},
\end{align}
where $\eta_{sr}$ represents the optical conversion ratio, $\daleth_{sr}$ indicates the photoelectric conversion factor, $I_{sr}$ defines the optical irradiance, and $\vartheta _{sr}$ denotes the additive white Gaussian noise (AWGN). Since $\mathcal{R}$ harvests RF energy from $\mathcal{B}$, the harvested energy at $\mathcal{R}$ is expressed as
\begin{align}
   \label{eqn61} 
   E_r=\eta_{r}  P_b \left| \mathbf{h}_{br}\right| ^2 \frac{T}{2},
\end{align}
where $P_b$ defines the transmit power at $\mathcal{B}$, $\mathbf{h}_{br}\in\mathbb{C}^{1\times G_{b}}$ is the channel gain of $\mathcal{B-R}$ link, and the energy conversion efficiency is indicated by $\eta_{r} (0\leq \eta_{r} \leq 1)$, which is mainly controlled by the harvesting circuitry.

In the second time frame, $\mathcal{R}$ firstly decodes the received optical signals and then transforms it into the RF domain. To facilitate the information transfer from $\mathcal{R}$ to $\mathcal{D}$, $\mathcal{R}$ utilizes all of the harvested energy with an output power of
\begin{align}
    \label{eqn62}
     P_{r}=\eta_{r}  P_b \left| \mathbf{h}_{br}\right| ^2.
\end{align}
The signals received at $\mathcal{D}$ is given as
\begin{align}
    \label{eqn63}
   \mathbf{y}_d= \sqrt{P_{r}} \mathbf{h}_{rd} z +\mathbf{n}_d,
\end{align}
where $z$ represents the transmitted signal, $\mathbf{h}_{rd}\in \mathbb{C}^{G_{d}\times 1}$ is the channel gain of $\mathcal{R-D}$ link, and the zero-mean AWGN with a noise power, $N_d$, is represented by $\mathbf{n}_d\sim \mathcal{\widetilde{N}}(0,N_{d}\mathbf{I}_{G_{d}})$. Here, $\mathbf{I}_{N}$ is an identity matrix of order $N \times N$. We consider $\mathcal{E}$ trying to wiretap the sensitive information signals via the illicit $\mathcal{R}-\mathcal{E}$ link. Hence, the received signals at $j$th eavesdropper in the second time frame are given as
\begin{align}
    \label{eqn64}
   \mathbf{y}_{e,j}= \sqrt{P_{r}} \mathbf{h}_{re,j}z + \mathbf{n}_e,
\end{align}
where $\mathbf{h}_{re,j}\in\mathbb{C} ^{G_{e}\times 1}$ is the channel gain of $\mathcal{R}\rightarrow j$th ($j=1,2,\ldots,\mathcal{E}$) link and the AWGN with zero mean and noise power, $N_e$, is represented by $\mathbf{n}_e\sim \mathcal{\widetilde{N}}(0,N_{e}\mathbf{I}_{G_{e}})$. The instantaneous SNRs for the $\mathcal{R-D}$ and $\mathcal{R-E}$ links are, respectively, defined as
\begin{align}
    \label{eqn65}
    \gamma_{rd}&=\frac{P_{r} \left| \mathbf{h}_{rd}\right| ^2}{N_d}=\frac{\eta_{r}  P_b }{N_d}\left| \mathbf{h}_{br}\right| ^2 \left| \mathbf{h}_{rd}\right| ^2,
\\
    \label{eqn66}
    \gamma_{re,j}&=\frac{P_{r} \left| \mathbf{h}_{re,j}\right| ^2}{N_e}=\frac{\eta_{r}  P_b }{N_e}\left| \mathbf{h}_{br}\right| ^2 \left| \mathbf{h}_{re,j}\right| ^2.
\end{align}

\section{Channel Model}
In this section, the channel modeling of UOWC $\mathcal{(S-R)}$ and RF $(\mathcal{R-D, R-E, B-R})$ links are realized for further mathematical analysis.

\subsection{PDF and CDF of SNR for $\mathcal{S-R}$ Link}
The PDF of $\gamma_{sr}$ can be defined as \cite[Eq.~11]{9622195}
\begin{align}
\label{eqn9}
f_{\gamma_{sr}}(\gamma)=\sum _{i=1}^2{B_i}\gamma^{-1} G_{1,2}^{2,0}\left(Z_i \gamma^{V_i}\biggl|
\begin{array}{c}
 W_i \\
 S_i,K_i \\
\end{array}
\right),
\end{align}
where
$B_1=\frac{\omega \xi ^2}{\epsilon}$, $B_2=\frac{\xi ^2 (1-\omega )}{\epsilon \Gamma (a)}$, $Z_1=\frac{1}{\mho \gamma \psi _\epsilon^{1/\epsilon}}$, $Z_2=\frac{1}{\gamma^{c} \sigma^c \psi _\epsilon^{c/\epsilon}}$, $V_1=\frac{1}{\epsilon}$, $V_2=\frac{c}{\epsilon}$, $W_1=\xi ^2+1$, $W_2=\frac{\xi ^2}{c}+1$, $S_1=1$, $S_2=a$, $K_1=\xi ^2$, $K_2=\frac{\xi ^2}{c}$, $\psi _1=\Phi _{sr}$, $\psi _2=\frac{\Phi _{sr}}{2 \omega \mho ^2+\sigma^2 (1-\omega )\frac{\Gamma \left(a+\frac{2}{c}\right)}{\Gamma (a)}}$, the average SNR of $\mathcal{S-R}$ link is indicated by $\Phi _{sr}$, $\mathcal{S-R}$ link's electrical SNR is defined by $\psi _\epsilon$, and $\epsilon$ represents the detection technique (e.g., $\epsilon = 1$ indicates HD technique and $\epsilon = 2$ implies IM/DD technique). The exponential distribution parameter is represented by $\mho$, three GG distributed constraints are symbolized by $a$, $\sigma$, and $c$, weight of the mixture is denoted by $0<\omega<1$, and pointing error is indicated by $\xi$. The values of $\omega$, $\mho$, $a$, $\sigma$, and $c$ are determined experimentally based on varying UWT (mild to severe) due to varying levels of air bubbles, temperature gradients, and water salinity. According to Table I in \cite{zedini2019unified}, increasing the temperature gradient and/or the level of air bubbles causes mild to severe UWT, which significantly raises the scintillation index. Table II in \cite{zedini2019unified} further displays several UWT scenarios in a thermally uniform UOWC network for fresh and salty waters.

The CDF of $\gamma_{sr}$ is given as \cite[Eq.~4]{sarker2020secrecy}
\begin{align}
\label{eqn10}
F_{\gamma_{sr}}(\gamma)=\int _0^\gamma f_{\gamma_{sr}}(\gamma)d\gamma.
\end{align}
Substituting \eqref{eqn9} into \eqref{eqn10}, $F_{\gamma_{sr}}(\gamma)$ is obtained as \cite[Eq.~12]{9622195}
\begin{align}
\label{eqn24}
F_{\gamma_{sr}}(\gamma)=\sum _{i=1}^2 Y_i G_{2,3}^{2,1}\left(Z_i \gamma^{V_i}\biggl|
\begin{array}{c}
 1,W_i \\
 S_i,K_i,0 \\
\end{array}
\right),
\end{align}
where $Y_1=\omega \xi ^2$ and $Y_2=\frac{\xi ^2 (1-\omega )}{c \Gamma (a)}$.

\subsection{PDF and CDF of SNR for $\mathcal{R-D}$ Link}

The PDF of $\gamma_{rd}$ is defined as \cite[Eq.~4]{9615217}
\begin{align}
\label{eqn200}
f_{\gamma_{rd}}(\gamma)= \alpha _1 e^{-\Xi _2 \gamma}  \gamma^{{\mu _d} -1} \, _1F_1\left(m_d,\mu _d;\alpha _2 \gamma \right),
\end{align}
where $\alpha _1=\frac{\left(G_d \mu _d\right)^{G_d \mu _d} \left(G_d m_d\right)^{G_d m_d}\left(1+\kappa _d\right)^{G_d \mu _d}}{\Gamma \left(G_d \mu _d\right)(\Phi _{rd})^{G_d \mu_d} \left(G_d \mu _d \kappa _d+G_d m_d\right)^{G_d m_d}}$, $\Xi _2=\frac{G_d \mu _d\left(1+\kappa _d\right)}{\Phi_{rd}}$, $\alpha _2=\frac{G_d^2\mu _d^2\kappa _d \left(1+\kappa _d\right)}{\left(G_d\mu _d\kappa _d+G_d m_d\right)\Phi _{rd}}$, the average SNR of $\mathcal{R-D}$ link is denoted by $\Phi_{rd}$, $G_d$ indicates the number of antennas of each user, the ratio of the powers of the dominant and scattered components, the number of clusters, and the Nakagami-$m$ faded shadowing parameter are symbolized by $\kappa _d$, $\mu_d$, and $m_d$, respectively. Here, $\, _1F_1(.,.;.)$ represents the confluent hyper-geometric function that can be expressed as $\, _1F_1\left(l_1,l_2;l_3\right)= \frac{\Gamma \left(l_2\right)}{\Gamma \left(l_1\right)} \sum _{i_1=0}^{\infty } \frac{ \Gamma \left(i_1+l_1\right)l_3^{i_1}}{\Gamma \left(i_1+l_2\right)i_1!}$ \cite[Eq.~13]{8070981}. Finally, $f_{\gamma_{rd}}(\gamma)$ can be written as \cite[Eq.~5]{9615217}
\begin{align}
\label{eqn1}
f_{\gamma_{rd}}(\gamma)= \sum _{e_1=0}^\infty \Xi _1 e^{-\Xi _2 \gamma} \gamma^{\Xi _3},
\end{align}
where $\Xi _1=\alpha _1 \alpha _3$, $\alpha _3=\frac{\Gamma \left(G_d \mu _d\right)}{\Gamma \left(G_d m_d\right)} \frac{\Gamma \left(G_d m_d+e_1\right)\alpha _2^{e_1}}{\Gamma \left(G_d \mu _d+e_1\right)e_1!}$, and $\Xi _3=G_d \mu _d-1+e_1$.
Similar to \eqref{eqn10}, utilizing the formula  \cite[Eq.~3.351.1]{gradshteyn2014table}, the CDF of $\gamma_{rd}$ is formulated as
\begin{align}
F_{\gamma_{rd}}(\gamma)=\sum _{e_1=0}^\infty \Xi _1\Biggl(\frac{\Xi _3!}{\Xi _2^{\Xi _3+1}}-\sum_{e_2=0}^{\Xi _3}\frac{\Xi _3!}{e_2!\Xi _2^{\Xi _3-e_2+1}} e^{-\Xi _2 \gamma} \gamma^{e_2}\Biggl).
\end{align}

\subsection{PDF and CDF of SNR for $\mathcal{R-E}$ Link}

The PDF of $\gamma_{re,j}$ can be expressed as \cite[Eq.~5]{9615217}
\begin{align}
\label{eqn4}
f_{\gamma_{re,j}}(\gamma)= \sum _{f_1=0}^{\infty} \Xi _4 e^{-\Xi _5 \gamma} \gamma^{\Xi _6},
\end{align}
where $\Xi _4=\beta _1 \beta _3$, $\Xi _5=\frac{G_e\mu _e \left(1+\kappa _e\right) }{\Phi _{re}}$, $\Xi _6=G_e \mu _e-1+f_1$, $\beta _1=\frac{\left(G_e \mu _e\right)^{G_e \mu _e} \left(G_e m_e\right)^{G_e m_e} \left(1+\kappa _e\right)^{G_e \mu _e}}{\Gamma \left(G_e \mu _e\right)(\Phi _{re})^{G_e \mu _e} \left(G_e\mu _e\kappa _e +G_e m_e\right)^{G_e m_e}}$, $\beta _2=\frac{G_e^2\mu _e^2\kappa _e \left(1+\kappa _e\right)}{\left(G_e\mu _e\kappa _e+G_e m_e\right)\Phi _{re}}$, $\beta _3=\frac{\Gamma \left(G_e \mu _e\right)}{\Gamma \left(G_e m_e\right)} \frac{\Gamma \left(G_e m_e+f_1\right)\beta _2^{f_1}}{\Gamma \left(G_e \mu _e+f_1\right)f_1!}$, $\Phi _{re}$ is the average SNR of $\mathcal{R-E}$ link, $G_e$ indicates the number of antennas of each eavesdropper, and the fading and shadowing parameters regarding $\mathcal{R-E}$ link are indicated by $\kappa _e$, $\mu _e$, and $m_e$. The CDF of $\gamma_{re,j}$ can be derived from \eqref{eqn4} making use of \cite[Eq.~3.351.2]{gradshteyn2014table} as
\begin{align}
\label{eqn14}
F_{\gamma_{re,j}}(\gamma)=1-\sum _{f_1=0}^{\infty} \sum _{p_1=0}^{\Xi _6} \frac{\Xi _6!}{p_1! \Xi _5^{\Xi _6-p_1+1}} \Xi _4 e^{-\Xi _5 \gamma} \gamma^{p_1}.
\end{align}

\subsubsection{Scenario-I}

For the colluding eavesdroppers, we substitute $\Phi_{re}$, $G_e$, $\mu_e$, and $m_e$ with $\mathcal{E}_I \Phi_{re}$, $\mathcal{E}_I G_e$, $\mathcal{E}_I \mu_e$, and $\mathcal{E}_I m_e$, respectively, based on the MRC technique \cite{9445659}. Here, $\mathcal{E}_I$ indicates the number of colluding eavesdroppers. For this scenario, the PDF of the instantaneous SNR of $\mathcal{R-E}$ link denoted by $\gamma_{re}$ is obtained as
\begin{align}
    \label{eqn70}
    f_{\gamma_{re}}^{I}(\gamma)= \sum _{f_2=0}^{\infty} \widetilde{\Xi _4} e^{-\widetilde{\Xi _5} \gamma} \gamma^{\widetilde{\Xi _6}},
\end{align}
where $\widetilde{\Xi _4}=\tilde{\beta _1}\tilde{\beta _3}$, $\widetilde{\Xi _5}=\frac{\mathcal{E}_I G_e\mu _e \left(1+\kappa _e\right) }{\mathcal{E}_I \Phi _{re}}$, $\widetilde{\Xi _6}= \mathcal{E}_I G_e \mu _e-1+f_2$, $\tilde{\beta _1}=\frac{\left(\mathcal{E}_I G_e \mu _e\right)^{\mathcal{E}_I G_e \mu _e} \left(\mathcal{E}_I G_e m_e\right)^{\mathcal{E}_I G_e m_e} \left(1+\kappa _e\right)^{\mathcal{E}_I G_e \mu _e}}{\Gamma \left(\mathcal{E}_I G_e \mu _e\right)(\mathcal{E}_I \Phi _{re})^{\mathcal{E}_I G_e \mu _e} \left(\mathcal{E}_I G_e\mu _e\kappa _e +\mathcal{E}_I G_e m_e\right)^{\mathcal{E}_I G_e m_e}}$, $\tilde{\beta _2}=\frac{(\mathcal{E}_I G_e)^2\mu _e^2\kappa _e \left(1+\kappa _e\right)}{\left(\mathcal{E}_I G_e\mu _e\kappa _e+\mathcal{E}_I G_e m_e\right)\mathcal{E}_I \Phi _{re}}$, and $\tilde{\beta _3}=\frac{\Gamma \left(\mathcal{E}_I G_e \mu _e\right)}{\Gamma \left(\mathcal{E}_I G_e m_e\right)} \frac{\Gamma \left(\mathcal{E}_I G_e m_e+f_2\right)(\tilde{\beta _2})^{f_2}}{\Gamma \left(\mathcal{E}_I G_e \mu _e+f_2\right)f_2!}$.

\subsubsection{Scenario-II}

In the scenario of non-colluding eavesdroppers, the PDF of instantaneous SNR is obtained utilizing the \textit{max} algorithm of order statistics as \cite[Eq.~4]{odeyemi2018physical}
\begin{align}
    \label{eqn71}
    f_{\gamma_{re}}^{II}(\gamma)=\mathcal{E}_{II} \biggl[F_{\gamma_{re,J}}(\gamma)\biggl]^{\mathcal{E}_{II}-1}f_{\gamma_{re,J}}(\gamma),
\end{align}
where $\mathcal{E}_{II}$ indicates the number of non-colluding eavesdroppers. Substituting \eqref{eqn4} and \eqref{eqn14} into \eqref{eqn71}, $f_{\gamma_{re}}^{II}(\gamma)$ is outlined as
\begin{align}
   \label{eqn202}
   \nonumber
   f_{\gamma_{re}}^{II}(\gamma)&=\mathcal{E}_{II}\Biggl[1-\sum _{f_1=0}^{\infty} \sum _{p_1=0}^{\Xi _6} \frac{\Xi _6!}{p_1! \Xi _5^{\Xi _6-p_1+1}} \Xi _4 e^{-\Xi _5 \gamma} \gamma^{p_1}\Biggl]^{\mathcal{E}_{II} -1}
   \\
   &\times \sum _{f_1=0}^{\infty} \Xi _4 e^{-\Xi _5 \gamma} \gamma^{\Xi _6}.
\end{align}
Utilizing the binomial theorem \cite[Eq.~1.111]{gradshteyn2014table} and performing some mathematical manipulations, equation \eqref{eqn202} is formulated as
\begin{align}
  \label{eqn203} 
  \nonumber
  f_{\gamma_{re}}^{II}(\gamma)&= \mathcal{E}_{II} \sum _{f_1=0}^{\infty} \Xi _4 e^{-\Xi _5 \gamma} \gamma^{\Xi _6} \Biggl[\sum _{h_1=0}^{\mathcal{E}_{II}-1}\binom{\mathcal{E}_{II}-1}{h_1}(-1)^{h_1}
  \\
  \nonumber
  & \times \Biggl(\sum _{f_1=0}^{\infty} \sum _{p_1=0}^{\Xi _6} \frac{\Xi _6!}{p_1! \Xi _5^{\Xi _6-p_1+1}} \Xi _4 e^{-\Xi _5 \gamma} \gamma^{p_1}\Biggl)^{h_1}\Biggl]
  \\
  \nonumber
  &=\mathcal{E}_{II} \sum _{f_1=0}^{\infty} \Xi _4 e^{-\Xi _5 \gamma} \gamma^{\Xi _6}\Biggl[\sum _{h_1=0}^{\mathcal{E}_{II}-1}\binom{\mathcal{E}_{II}-1}{h_1}(-1)^{h_1}
  \\
  &\times \Biggl(\sum _{f_1=0}^{\infty}\Xi _4 e^{-\Xi _5\gamma}\Biggl)^{h_1}\Biggl(\sum _{p_1=0}^{\Xi _6} \frac{\Xi _6!}{p_1!}\frac{\gamma^{p_1}}{\Xi _5^{\Xi _6-p_1+1}}\Biggl)^{h_1}\Biggl].
\end{align}
Since $\biggl(\sum _{p_1=0}^{\Xi _6} \frac{\Xi _6!}{p_1!}\frac{\gamma^{p_1}}{\Xi _5^{\Xi _6-p_1+1}}\biggl)^{h_1}$ portion of equation \eqref{eqn203} is normally difficult to solve, we use the multinomial theorem \cite{BADRUDDUZA2020101177} to it. Applying the multinomial theorem, we write $\biggl(\sum _{p_1=0}^{\Xi _6} \frac{\Xi _6!}{p_1!}\frac{\gamma^{p_1}}{\Xi _5^{\Xi _6-p_1+1}}\biggl)^{h_1}$ portion as
\begin{align}
   \label{eqn210} 
   \nonumber
    \Biggl(\sum _{p_1=0}^{\Xi _6}  \frac{\Xi _6!}{p_1!} \frac{\gamma^{p_1}}{\Xi _5^{\Xi _6-p_1+1}}\Biggl)^{h_1}&=\sum _{q_0+q_1+\ldots+q_{\Xi _6}=h_1} \binom{h_1}{ q_0,q_1,\ldots,q_{\Xi _6}} 
   \\
   & \times \prod _{p_1} \biggl(\frac{\Xi _6!}{p_1!\Xi _5^{\Xi _6-p_1+1}}\biggl)^{q_{p_1}} \gamma^{\sum _{p_1} p_1 q_{p_1}}.
\end{align}
Substituting \eqref{eqn210} in equation \eqref{eqn203} and performing some mathematical manipulations, $f_{\gamma_{re}}^{II}(\gamma)$ is finally derived as
\begin{align}
  \label{eqn204}  
  \nonumber
  f_{\gamma_{re}}^{II}(\gamma)&= \mathcal{E}_{II} \sum _{f_1=0}^{\infty} \sum _{h_1=0}^{\mathcal{E}_{II}-1} \sum _{q_0+q_1+\ldots+q_{\Xi _6}=h_1} \prod _{p_1} \binom{\mathcal{E}_{II}-1}{h_1}
  \\
  \nonumber
  &\times \binom{h_1}{ q_0,q_1,\ldots,q_{\Xi _6}} \Biggl(\frac{\Xi _6!}{p_1!\Xi _5^{\Xi _6-p_1+1}}\Biggl)^{q_{p_1}}
  \\
  & \times \Biggl(\sum _{f_1=0}^{\infty}\Xi _4\Biggl)^{h_{1}} (-1)^{h_1} \Xi _4 e^{-\gamma(\Xi _5+\Xi _5 h_1)}   \gamma^{\mathcal{X} _3},
\end{align}
where $\mathcal{X} _3= \sum _{p_1} p_1 q_{p_1}+\Xi _6$.
\subsection{PDF and CDF of SNR for $\mathcal{B-R}$ Link}

The PDF of $\gamma_{br}$ can be addressed as \cite[Eq.~5]{9615217}
\begin{align}
\label{eqn7}
f_{\gamma_{br}}(\gamma)= \sum _{g_1=0}^\infty \Xi _7 e^{-\Xi _8 \gamma} \gamma^{\Xi _9},
\end{align}
where $\Xi _7=\lambda _1 \lambda _3$, $\Xi _8=\frac{G_b \mu _b\left(1+\kappa _b\right)}{\Phi_{br}}$, $\Xi _9=G_b \mu _b-1+g_1$, $\lambda _1=\frac{\left(G_b \mu _b\right)^{G_b \mu _b} \left(G_b m_b\right)^{G_b m_b}\left(1+\kappa _b\right)^{G_b \mu _b}}{\Gamma \left(G_b \mu _b\right)(\Phi _{br})^{G_b \mu_b} \left(G_b \mu _b \kappa _b+G_b m_b\right)^{G_b m_b}}$, $\lambda _2=\frac{G_b^2\mu _b^2\kappa _b \left(1+\kappa _b\right)}{\left(G_b\mu _b\kappa _b+G_b m_b\right)\Phi _{br}}$, $\lambda _3=\frac{\Gamma \left(G_b \mu _b\right)}{\Gamma \left(G_b m_b\right)} \frac{\Gamma \left(G_b m_b+g_1\right)\lambda _2^{g_1}}{\Gamma \left(G_b \mu _b+g_1\right)g_1!}$, the average SNR of $\mathcal{B-R}$ link is defined by $\Phi _{br}$, $G_b$ indicates the number of antennas of the beacon, and similar to $\mathcal{R-D}$ link, the channel parameters corresponding to $\mathcal{B-R}$ link are denoted by $\kappa _b$, $\mu _b$, and $m_b$. The CDF of $\gamma_{br}$ is expressed as
\begin{align}
F_{\gamma_{br}}(\gamma)=\sum _{g_1=0}^\infty \Xi _7\Biggl(\frac{\Xi _9!}{\Xi _8^{\Xi _9+1}}-\sum_{g_2=0}^{\Xi _9}\frac{\Xi _9!}{g_2!\Xi _8^{\Xi _9-g_2+1}} e^{-\Xi _8 \gamma} \gamma^{g_2}\Biggl).
\end{align}

\section{Performance Metrics}

SOP, SPSC, and EST are three key performance parameters that are frequently utilized to evaluate physical layer secrecy performance. This section demonstrates closed-form expressions for those performance metrics.

\subsection{Secrecy Outage Probability Analysis}

For the first and second hops, the instantaneous secrecy capacities (SC) are defined, respectively, as
\begin{align}
\label{eqn11}
C_{sr}&=\frac{1}{2}\log _2 \left(1+\gamma_{sr}\right),
\\
\label{eqn12}
C_{rd}&=\left\{\frac{1}{2} \left[\log _2 \left(1+\gamma_{rd}\right)-\log _2 \left(1+\gamma_{re,j}\right)\right]\right\}^+,
\end{align}
where $\{z\}^+=\max (z,0)$. Furthermore, the well-known \textit{max-flow min-cut} theory states that the system's instantaneous capability ($C_m$) is restricted by the nominal capacity of the two hops and it is defined as \cite[Eq.~14]{pan2019secrecy}
\begin{align}
\label{eqn13}
C_m=\min \left(C_{sr},C_{rd}\right).
\end{align}
The SOP is the probability of $C_m$ falling below a target secrecy rate, $R_s$ $(R_s>0)$, expressed as
\begin{align}
\label{eqn15}
\nonumber
{\text{SOP}}&=\Pr \left\{C_m<R_s\right\}=\Pr\left\{\min \left(C_{sr},C_{rd}\right)<R_s\right\}
\\
\nonumber
&=1-\Pr\left\{\min \left(C_{sr},C_{rd}\right)\geq R_s\right\}
\\
&=1-\Pr \left\{C_{sr}\geq R_s\right\}\Pr \left\{C_{rd}\geq R_s\right\},
\end{align}
wherein
\begin{align}
\label{eqn16}
\nonumber
\Pr \left\{C_{sr}\geq R_s\right\}&=1-\Pr \left\{C_{sr}<R_s\right\}
\\
\nonumber
&=1-\Pr \left\{\frac{1}{2} \log _2 \left(1+\gamma_{sr}\right)<R_s\right\}
\\
&=1-\Pr \left\{\gamma_{sr}<\theta -1\right\}=1- F_{\gamma_{sr}}(\theta -1),
\end{align}
where $\theta =2^{2 R_s}$. Substituting \eqref{eqn24} into \eqref{eqn16}, $\Pr \left\{C_{sr}\geq R_s\right\}$ is formulated as
\begin{align}
\label{eqn25}
\Pr \left\{C_{sr}\geq R_s\right\}=1-\sum _{i=1}^2 Y_i G_{2,3}^{2,1}\left(Z_i (\theta -1)^{V_i}\biggl|
\begin{array}{c}
 1,W_i \\
 S_i,K_i,0 \\
\end{array}
\right),
\end{align}
and
\begin{align}
\label{eqn17}
\nonumber
\Pr &\left\{C_{rd}\geq R_s\right\}
\\
\nonumber
&=\Pr \left\{\frac{1}{2}\left[ \log _2 \left(\gamma_{rd}+1\right)-\log _2 \left(\gamma_{re,j}+1\right)\right]\geq R_s\right\}
\\
\nonumber
&=\Pr \left\{\gamma_{rd}\geq \theta  \gamma_{re,j}\right(\theta -1)\}
\\
\nonumber
&=\Pr \left\{\frac{\eta_{r}  P_b \left| \mathbf{h}_{br}\right|^2}{N_d} \left(\left| \mathbf{h}_{rd}\right|^2-\theta  \left| \mathbf{h}_{re,j}\right| ^2\right) \geq \theta -1\right\}
\\
&=\Pr \left\{\left| \mathbf{h}_{br}\right| ^2 w \geq D_0\right\},
\end{align}
where $D_0=\frac{(\theta -1) N_d}{\eta_{r}  P_b}$ and $w=\left| \mathbf{h}_{rd}\right| ^2-\theta  \left| \mathbf{h}_{re,j}\right| ^2$. Here, the term $N_d$ is set to $1$. To hold the inequality $ \left| \mathbf{h}_{br}\right| ^2 w\geq D_0$, it has $w>0$. Hence, we obtain
\begin{align}
\label{eqn18}
\nonumber
\Pr \left\{C_{rd}\geq R_s\right\}&= \Pr \left\{\left| \mathbf{h}_{br}\right| ^2 w\geq D_0,w>0\right\}
\\
&= \int _0^{\infty }\int _{\frac{D_0}{w}}^{\infty } f_{\gamma_{br}}(\gamma) f_w (w) d\gamma dw,
\end{align}
whereby
\begin{align}
\label{eqn19}
f_w(w)=\int _0^{\infty } f_{\gamma_{rd}}(w+x) \frac{1}{\theta }f_{\gamma_{re,j}}\left(\frac{x}{\theta }\right)dx.
\end{align}

\subsubsection{Scenario-I}

In the scenario of colluding eavesdroppers, equation \eqref{eqn18} is expressed as
\begin{align}
\label{eqn100}
\Pr \left\{\left| \mathbf{h}_{br}\right| ^2 w\geq D_0,w>0\right\} &= \int _0^{\infty }\int _{\frac{D_0}{w}}^{\infty } f_{\gamma_{br}}(\gamma)f_w^{I}(w) d\gamma dw,
\end{align}
wherein
\begin{align}
\label{eqn74}
 f_w^{I}(w)=\int _0^{\infty } f_{\gamma_{rd}}(w+x) \frac{1}{\theta }f_{\gamma_{re}}^{I}\left(\frac{x}{\theta }\right)dx.
\end{align}
Placing \eqref{eqn1} and \eqref{eqn70} into \eqref{eqn74}, $f_w^{I}(w)$ is expressed as
\begin{align}
\label{eqn26}
\nonumber
f_w^{I}(w)&= \int _0^{\infty } \sum _{e_1=0}^\infty \Xi _1 e^{-\Xi _2 (w+x)} (w+x)^{\Xi _3} \frac{1}{\theta }\sum _{f_2=0}^\infty \widetilde{\Xi _4}
\\
& \times e^{-\widetilde{\Xi _5} (\frac{x}{\theta })} \left(\frac{x}{\theta }\right)^{\widetilde{\Xi _6}}dx.
\end{align}
Applying the binomial theorem to $(w+x)^{\Xi _3}$ term, utilizing the formula \cite[Eq.~3.351.3]{gradshteyn2014table} along with some mathematical manipulations and simplifications, equation \eqref{eqn26} is derived as
\begin{align}
\label{eqn215}
\nonumber
f_w^{I}(w)&= \sum _{e_1=0}^\infty \sum _{f_2=0}^\infty \sum _{t_1=0}^{\Xi _3} \binom{\Xi _3}{t_1}\frac{\mathcal{X} _1}{\theta ^{\widetilde{\Xi _6}+1}} e^{-\Xi _2 w} w^{\Xi _3-t_1}
\\
\nonumber
& \times \int _0^{\infty } e^{-(\Xi _2+\frac{\widetilde{\Xi _5}}{\theta })x} x^{\widetilde{\Xi _6}+t_1}dx
\\
\nonumber
&= \sum _{e_1=0}^\infty \sum _{f_2=0}^\infty \sum _{t_1=0}^{\Xi _3} \binom{\Xi _3}{t_1} \biggl(\Xi _2+\frac{\widetilde{\Xi _5}}{\theta }\biggl)^{-(\widetilde{\Xi _6}+t_1+1)} 
\\
& \times \frac{(\widetilde{\Xi _6}+t_1)! \mathcal{X} _1}{\theta ^{\widetilde{\Xi _6}+1}}  e^{-\Xi _2 w} w^{\Xi _3-t_1},
\end{align}
where $\mathcal{X} _1=\Xi _1 \widetilde{\Xi _4}$. Now, substituting \eqref{eqn7} and \eqref{eqn215} into \eqref{eqn100}, utilizing the formula \cite[Eq.~3.351.2]{gradshteyn2014table}, and performing some mathematical manipulations and simplifications, $\Pr \left\{\left| \mathbf{h}_{br}\right| ^2 w\geq D_0,w>0\right\}$ is formulated as
\begin{align}
\label{eqn27}
\nonumber
\Pr &\left\{\left| \mathbf{h}_{br}\right| ^2 w\geq D_0,w>0\right\}
\\
\nonumber
&=\sum _{e_1=0}^\infty \sum _{f_2=0}^\infty\sum _{g_1=0}^\infty \sum _{t_1=0}^{\Xi _3} \binom{\Xi _3}{t_1} \biggl(\Xi _2+\frac{\widetilde{\Xi _5}}{\theta }\biggl)^{-(\widetilde{\Xi _6}+t_1+1)} 
\\
\nonumber
& \times \frac{(\widetilde{\Xi _6}+t_1)! \mathcal{X} _2}{\theta ^{\widetilde{\Xi _6}+1}} \int _0^{\infty }e^{-\Xi _2 w} w^{\Xi _3-t_1}dw \int _{\frac{D_0}{w}}^{\infty }e^{-\Xi _8 \gamma} \gamma^{\Xi _9}d\gamma
\\
\nonumber
&=\sum _{e_1=0}^\infty \sum _{f_2=0}^\infty\sum _{g_1=0}^\infty \sum _{t_1=0}^{\Xi _3} \sum _{t_2=0}^{\Xi _9} \binom{\Xi _3}{t_1}\biggl(\Xi _2+\frac{\widetilde{\Xi _5}}{\theta }\biggl)^{-(\widetilde{\Xi _6}+t_1+1)} 
\\
\nonumber
& \times \frac{(\widetilde{\Xi _6}+t_1)! \mathcal{X}_2}{\theta ^{\widetilde{\Xi _6}+1}} \frac{\Xi _9!}{t_2!\Xi _8^{\Xi _9-t_2+1}} \int _0^{\infty }\left(\frac{D_0}{w}\right)^{t_2}
\\
& \times  e^{-\Xi _2 w - \frac{\Xi _8 D_0}{w}} w^{\Xi _3-t_1} dw,
\end{align}
where $\mathcal{X} _2=\mathcal{X} _1 \Xi _7$. Applying the formula \cite[Eq.~3.471.9]{gradshteyn2014table} after performing some mathematical simplifications, equation \eqref{eqn27} is finally implemented as
\begin{align}
\label{eqn28}
\nonumber
\Pr &\left\{\left| \mathbf{h}_{br}\right| ^2 w\geq D_0,w>0\right\}
\\
\nonumber
&= 2\sum _{e_1=0}^\infty \sum _{f_2=0}^\infty\sum _{g_1=0}^\infty \sum _{t_1=0}^{\Xi _3} \sum _{t_2=0}^{\Xi _9} \binom{\Xi _3}{t_1} \biggl(\Xi _2+\frac{\widetilde{\Xi _5}}{\theta }\biggl)^{-(\widetilde{\Xi _6}+t_1+1)}
\\
\nonumber
& \times \frac{(\widetilde{\Xi _6}+t_1)! \mathcal{X}_2}{\theta ^{\widetilde{\Xi _6}+1}} \frac{\Xi _9! D_0^{t_2}}{t_2!\Xi _8^{\Xi _9-t_2+1}} \left(\frac{\Xi _8 D_0}{\Xi _2}\right)^{\frac{\Xi _3-t_1-t_2+1}{2}}
\\
& \times K_{\Xi _3-t_1-t_2+1}\left(2 \sqrt{\Xi _2 \Xi _8 D_0}\right),
\end{align}
where $K_{v}(.)$ is modified Bessel function of second kind. Substituting \eqref{eqn25} and \eqref{eqn28} into \eqref{eqn15}, we finally obtain the SOP expression for Scenario-I that is presented in \eqref{eqn29}.
\begin{figure*}[!b]
\vspace{0mm}
\hrulefill
\begin{align}
\nonumber
\label{eqn29}
\text{SOP}^{I}&=1-\Biggl[2\sum _{e_1=0}^\infty \sum _{f_2=0}^\infty\sum _{g_1=0}^\infty \sum _{t_1=0}^{\Xi _3} \sum _{t_2=0}^{\Xi _9} \binom{\Xi _3}{t_1} \biggl(\Xi _2+\frac{\widetilde{\Xi _5}}{\theta }\biggl)^{-(\widetilde{\Xi _6}+t_1+1)} \frac{(\widetilde{\Xi _6}+t_1)! \mathcal{X}_2}{\theta ^{\widetilde{\Xi _6}+1}} \frac{\Xi _9! D_0^{t_2}}{t_2!\Xi _8^{\Xi _9-t_2+1}} \left(\frac{\Xi _8 D_0}{\Xi _2}\right)^{\frac{\Xi _3-t_1-t_2+1}{2}} 
\\
& \times K_{\Xi _3-t_1-t_2+1}\left(2 \sqrt{\Xi _2 \Xi _8 D_0}\right)\Biggl]\left[1-\sum _{i=1}^2 Y_i G_{2,3}^{2,1}\left(Z_i (\theta -1)^{V_i}\biggl|
\begin{array}{c}
 1,W_i \\
 S_i,K_i,0 \\
\end{array}
\right)\right]
\end{align}
\vspace{0mm}
\end{figure*}
\subsubsection{Scenario-II}
In the scenario with non-colluding eavesdroppers, equation \eqref{eqn18} is expressed as
\begin{align}
\label{eqn101}
\Pr \left\{\left| \mathbf{h}_{br}\right| ^2 w\geq D_0,w>0\right\} &= \int _0^{\infty }\int _{\frac{D_0}{w}}^{\infty } f_{\gamma_{br}}(\gamma)f_w^{II}(w) d\gamma dw,
\end{align}
where
\begin{align}
\label{eqn45}
f_w^{II}(w)=\int _0^{\infty } f_{\gamma_{rd}}(w+x) \frac{1}{\theta }f_{\gamma_{re}}^{II}\left(\frac{x}{\theta }\right)dx.
\end{align}
Substituting \eqref{eqn1} and \eqref{eqn204} into \eqref{eqn45}, $f_w^{II}(w)$ is expressed as
\begin{align}
     \label{eqn220}
     \nonumber
     f_w^{II}(w)&=\int _0^{\infty } \sum _{e_1=0}^\infty \Xi _1 e^{-\Xi _2 (w+x)} (w+x)^{\Xi _3} \frac{1}{\theta}\mathcal{E}_{II}
     \\
     \nonumber
     & \times  \sum _{f_1=0}^{\infty} \sum _{h_1=0}^{\mathcal{E}_{II}-1} \sum _{q_0+q_1+\ldots+q_{\Xi _6}=h_1} \prod _{p_1} \binom{\mathcal{E}_{II}-1}{h_1}
     \\
     \nonumber
     &\times \binom{h_1}{ q_0,q_1,\ldots,q_{\Xi _6}} \Biggl(\frac{\Xi _6!}{p_1!\Xi _5^{\Xi _6-p_1+1}}\Biggl)^{q_{p_1}}
     \\
     & \times \Biggl(\sum _{f_1=0}^{\infty}\Xi _4\Biggl)^{h_{1}} (-1)^{h_1} \Xi _4 e^{-\frac{x}{\theta }(\Xi _5+\Xi _5 h_1)}   \left(\frac{x}{\theta }\right)^{\mathcal{X} _3}.
\end{align}
Now, according to the similar formulation procedure as utilized for scenario-I with some mathematical manipulations and simplifications, $f_w^{II}(w)$ is finally derived that is shown in \eqref{eqn49}, where $\mathcal{X} _4= \mathcal{X} _3 + t_3$.
\begin{figure*}[!b]
\vspace{0mm}
\hrulefill
\begin{align}
\label{eqn49}
\nonumber
f_w^{II}(w)&= \mathcal{E}_{II}\sum _{e_1=0}^\infty \sum _{f_1=0}^\infty \sum _{h_1=0}^{\mathcal{E}_{II}-1}\sum _{q_0+q_1+\ldots+q_{\Xi _6}=h_1}\sum _{t_3=0}^{\Xi _3} \prod _{p_1} \binom{\mathcal{E}_{II}-1}{h_1}\binom{\Xi _3}{t_3} \binom{h_1}{ q_0,q_1,\ldots,q_{\Xi _6}} \Biggl(\frac{\Xi _6!}{p_1!\Xi _5^{\Xi _6-p_1+1}}\Biggl)^{q_{p_1}}
\\
&\times \Biggl(\sum _{f_1=0}^{\infty}\Xi _4\Biggl)^{h_{1}} \left(\Xi _2+\frac{\Xi _5 h_1}{\theta }+\frac{\Xi _5}{\theta }\right)^{-(\mathcal{X} _4+1)}  (-1)^{h_1} \frac{\mathcal{X} _4!}{\theta ^{\mathcal{X} _3+1}}\Xi _1 \Xi _4  e^{-\Xi _2 w}w^{\Xi _3-t_1}
\end{align}
\vspace{0mm}
\end{figure*}

Substituting \eqref{eqn7} and \eqref{eqn49} into \eqref{eqn101}, implementing the similar formulation approach as utilized for scenario-I along with some mathematical manipulations and simplifications, $\Pr \left\{\left| \mathbf{h}_{br}\right| ^2 w\geq D_0,w>0\right\}$ is obtained as shown in \eqref{eqn50}, where $\mathcal{X} _5=\Xi _1 \Xi _4 \Xi _7$.
\begin{figure*}[!b]
\vspace{0mm}
\hrulefill
\begin{align}
   \label{eqn50}
   \nonumber
   \Pr \left\{\left| \mathbf{h}_{br}\right| ^2 w\geq D_0,w>0\right\}&= 2\mathcal{E}_{II}\sum _{e_1=0}^\infty \sum _{f_1=0}^\infty \sum _{g_1=0}^\infty \sum _{h_1=0}^{\mathcal{E}_{II}-1}\sum _{t_3=0}^{\Xi _3} \sum _{t_4=0}^{\Xi _9} \sum _{q_0+q_1+\ldots+q_{\Xi _6}=h_1} \prod _{p_1} \binom{\mathcal{E}_{II}-1}{h_1}\binom{\Xi _3}{t_3} \binom{h_1}{ q_0,q_1,\ldots,q_{\Xi _6}}
  \\
  \nonumber
  & \times \Biggl(\frac{\Xi _6!}{p_1!\Xi _5^{\Xi _6-p_1+1}}\Biggl)^{q_{p_1}} \Biggl(\sum _{f_1=0}^{\infty}\Xi _4\Biggl)^{h_{1}} \left(\Xi _2+\frac{\Xi _5 h_1}{\theta }+\frac{\Xi _5}{\theta }\right)^{-(\mathcal{X} _4+1)}(-1)^{h_1} \frac{\mathcal{X} _4!\mathcal{X} _5}{\theta ^{\mathcal{X} _3+1}} \frac{\Xi _9! D_0^{t_4}}{t_4!\Xi _8^{\Xi _9-t_4+1}}
  \\
  & \times \left(\frac{\Xi _8 D_0}{\Xi _2}\right)^{\frac{\Xi _3-t_3-t_4+1}{2}}  K_{\Xi _3-t_3-t_4+1}\left(2 \sqrt{\Xi _2 \Xi _8 D_0}\right)
\end{align}
\vspace{0mm}
\end{figure*}

Finally, placing \eqref{eqn25} and \eqref{eqn50} into \eqref{eqn15}, we obtain the SOP formula for scenario-II which is expressed in \eqref{eqn52}.
\begin{figure*}[!b]
\vspace{0mm}
\hrulefill
\begin{align}
  \label{eqn52}  
  \nonumber
  \text{SOP}^{II}&=1-\Biggl[2\mathcal{E}_{II}\sum _{e_1=0}^\infty \sum _{f_1=0}^\infty \sum _{g_1=0}^\infty \sum _{h_1=0}^{\mathcal{E}_{II}-1}\sum _{t_3=0}^{\Xi _3} \sum _{t_4=0}^{\Xi _9} \sum _{q_0+q_1+\ldots+q_{\Xi _6}=h_1} \prod _{p_1} \binom{\mathcal{E}_{II}-1}{h_1}\binom{\Xi _3}{t_3} \binom{h_1}{ q_0,q_1,\ldots,q_{\Xi _6}}
  \\
  \nonumber
  & \times \Biggl(\frac{\Xi _6!}{p_1!\Xi _5^{\Xi _6-p_1+1}}\Biggl)^{q_{p_1}} \Biggl(\sum _{f_1=0}^{\infty}\Xi _4\Biggl)^{h_{1}} \left(\Xi _2+\frac{\Xi _5 h_1}{\theta }+\frac{\Xi _5}{\theta }\right)^{-(\mathcal{X} _4+1)}(-1)^{h_1} \frac{\mathcal{X} _4!\mathcal{X} _5}{\theta ^{\mathcal{X} _3+1}} \frac{\Xi _9! D_0^{t_4}}{t_4!\Xi _8^{\Xi _9-t_4+1}}
  \\
  & \times \left(\frac{\Xi _8 D_0}{\Xi _2}\right)^{\frac{\Xi _3-t_3-t_4+1}{2}}  K_{\Xi _3-t_3-t_4+1}\left(2 \sqrt{\Xi _2 \Xi _8 D_0}\right)\Biggl]\left[1-\sum _{i=1}^2 Y_i G_{2,3}^{2,1}\left(Z_i (\theta -1)^{V_i}\biggl|
\begin{array}{c}
 1,W_i \\
 S_i,K_i,0 \\
\end{array}
\right)\right]
\end{align}
\vspace{0mm}
\end{figure*}

\subsection{Strictly Positive Secrecy Capacity Analysis}

The probability of SPSC refers to the probability when $C_m >0$. SPSC is a significant aspect of the system's secrecy transmission. Although analytical representation of the probability of SPSC can be simply generated from the SOP formulation, it's physical significance is different from SOP. Hence, it can be defined as \cite{islam2021impact}
\begin{align}
\label{eqn40}
\nonumber
{\text{SPSC}}&=\Pr \left\{C_m>0\right\}=\Pr\left\{\min \left(C_{sr},C_{rd}\right)>0\right\}
\\
&=\Pr \left\{C_{sr}>0\right\}\Pr \left\{C_{rd}>0\right\},
\end{align}
where
\begin{align}
\label{eqn21}
\nonumber
\Pr \left\{C_{sr}>0\right\}&=\Pr \left\{\frac{1}{2} \log _2 \left(1+\gamma_{sr}\right)>0\right\}
\\
&=\Pr \left\{\gamma_{sr}>0\right\}=1,
\end{align}
and
\begin{align}
\label{eqn22}
\nonumber
\Pr &\left\{C_{rd}>0\right\}
\\
\nonumber
&=\Pr \left\{\frac{1}{2} \left[\log _2 \left(1+\gamma_{rd}\right)-\log _2 \left(1+\gamma_{re,j}\right)\right]>0\right\}
\\
\nonumber
&=1-\Pr \left\{\gamma_{rd}\leq \gamma_{re,j}\right\}=1-\Pr \left\{\left| \mathbf{h}_{rd}\right| ^2\leq \left| \mathbf{h}_{re,j}\right| ^2\right\}
\\
&=1-\int _0^{\infty }\int _0^v f_{\gamma_{rd}}(u) f_{\gamma_{re,j}}(v) du dv.
\end{align}

\subsubsection{Scenario-I}
For the scenario of colluding eavesdroppers, equation \eqref{eqn22} is expressed as
\setcounter{eqnback}{\value{equation}} \setcounter{equation}{48}
\begin{align}
    \label{eqn80}
    \Pr \left\{C_{rd}>0\right\}=1-\int _0^{\infty }\int _0^v f_{\gamma_{rd}}(u) f_{\gamma_{re}}^{I}(v) du dv.
\end{align}
Placing \eqref{eqn1} and \eqref{eqn70} into \eqref{eqn80}, utilizing the formula \cite[Eq.~3.351.1]{gradshteyn2014table}, and performing some mathematical manipulations and simplifications, $\Pr \left\{C_{rd}>0\right\}$ is formed as
\begin{align}
\label{eqn35}
\nonumber
& \Pr \left\{C_{rd}>0\right\}
\\
\nonumber
&=1-\Biggl[\sum _{e_1=0}^\infty \sum _{f_2=0}^\infty \mathcal{X}_1 \Biggl(\int _0^{\infty }e^{-\widetilde{\Xi _5} v} v^{\widetilde{\Xi _6}}dv \int _0^v e^{-\Xi _2 u}u^{\Xi _3} du \Biggl)\Biggl]
\\
\nonumber
&=1-\Biggl[\sum _{e_1=0}^\infty \sum _{f_2=0}^\infty \mathcal{X} _1 \Biggl( \int _0^{\infty } \frac{\Xi _3!}{\Xi _2^{\Xi _3+1}} e^{-\widetilde{\Xi _5} v} v^{\widetilde{\Xi _6}} dv 
\\
& -\int _0^{\infty } \sum _{t_5=0}^{\Xi _3} \frac{\Xi _3!} {t_5! \Xi _2^{\Xi _3-t_5+1}} e^{-(\Xi _2 + \widetilde{\Xi _5})v} v^{\widetilde{\Xi _6}+t_5} dv \Biggl) \Biggl],
\end{align}
where $\mathcal{X} _1=\Xi _1 \widetilde{\Xi _4}$. Now, applying the formula \cite[Eq.~3.351.3]{gradshteyn2014table}, $\Pr \left\{C_{rd}>0\right\}$ is finally formulated as
\begin{align}
\label{eqn41} 
\nonumber
\Pr \left \{C_{rd}>0\right\}&=1- \Biggl[\sum _{e_1=0}^\infty \sum _{f_2=0}^\infty \mathcal{X} _1\Biggl(\frac{\Xi _3! \widetilde{\Xi _6}!}{\Xi _2^{\Xi _3+1}} \widetilde{\Xi _5}^{-(\widetilde{\Xi _6}+1)}
\\
& -\sum _{t_5=0}^{\Xi _3} \frac{\Xi _3!(\widetilde{\Xi _6}+t_5)!}{t_5!\Xi _2^{\Xi _3-t_5+1}}(\Xi _2+\widetilde{\Xi _5})^{-(\widetilde{\Xi _6}+t_5+1)} \Biggl) \Biggl].
\end{align}
Placing \eqref{eqn21} and \eqref{eqn41} into \eqref{eqn40}, The formula of SPSC for the scenario-I is expressed as
\begin{align}
\label{eqn42}
\nonumber
\text{SPSC}^{I}&=1- \Biggl[\sum _{e_1=0}^\infty \sum _{f_2=0}^\infty \mathcal{X} _1\Biggl(\frac{\Xi _3! \widetilde{\Xi _6}!}{\Xi _2^{\Xi _3+1}} \widetilde{\Xi _5}^{-(\widetilde{\Xi _6}+1)}
\\
& -\sum _{t_5=0}^{\Xi _3} \frac{\Xi _3! (\widetilde{\Xi _6}+t_5)!}{t_5!\Xi _2^{\Xi _3-t_5+1}}(\Xi _2+\widetilde{\Xi _5})^{-(\widetilde{\Xi _6}+t_5+1)} \Biggl) \Biggl].
\end{align}

\subsubsection{Scenario-II}
In the scenario of non-colluding eavesdroppers, $\Pr \left\{C_{rd}>0\right\}$ is expressed as
\begin{align}
    \label{eqn53}
    \Pr &\left\{C_{rd}>0\right\}= 1-\int _0^{\infty }\int _0^v f_{\gamma_{rd}}(u) f_{\gamma_{re}}^{II}(v) du dv.
\end{align}
Substituting \eqref{eqn1} and \eqref{eqn204} into \eqref{eqn53}, following the similar formulation procedure as utilized for scenario-I with some mathematical manipulations and simplifications, $\Pr \left\{C_{rd}>0\right\}$ is derived as
\begin{align}
    \label{eqn225}
    \nonumber
    \Pr &\left\{C_{rd}>0\right\}
    \\
    \nonumber
    &= 1-\Biggl[\mathcal{E}_{II}\sum _{e_1=0}^\infty \sum _{f_1=0}^\infty \sum _{h_1=0}^{\mathcal{E}_{II}-1}\sum _{q_0+q_1+\ldots+q_{\Xi _6}=h_1} \prod _{p_1}  
    \\
    \nonumber
    & \times  \binom{\mathcal{E}_{II}-1}{h_1} \binom{h_1}{ q_0,q_1,\ldots,q_{\Xi _6}} \Biggl(\frac{\Xi _6!}{p_1!\Xi _5^{\Xi _6-p_1+1}}\Biggl)^{q_{p_1}}
    \\
    \nonumber
    & \times \Biggl(\sum _{f_1=0}^{\infty}\Xi _4\Biggl)^{h_{1}} (-1)^{h_1} \mathcal{X} _6 \Biggl ( \frac{\Xi _3! \mathcal{X} _3!}{\Xi _2^{\Xi _3+1}}  (\Xi _5+ \Xi _5 h_1)^{-(\mathcal{X} _3+1)}
    \\
    & - \sum _{t_6=0}^{\Xi _3} \frac{\Xi _3! \mathcal{X} _7!}{t_6!\Xi _2^{\Xi _3-t_6+1}} (\Xi _2+\Xi _5 h_1+\Xi _5)^{-(\mathcal{X} _7+1)}\Biggl)\Biggl],
\end{align}
where $\mathcal{X} _6= \Xi _1 \Xi _4$ and $\mathcal{X} _7= \mathcal{X} _3 + t_6$. Substituting \eqref{eqn21} and \eqref{eqn225} into \eqref{eqn40}, we can finally demonstrate the formula of SPSC for scenario-II as presented in \eqref{eqn58}.

\begin{figure*}[!b]
\vspace{0mm}
\hrulefill
\begin{align}
   \label{eqn58}
   \nonumber
   \text{SPSC}^{II}&=1-\Biggl[\mathcal{E}_{II}\sum _{e_1=0}^\infty \sum _{f_1=0}^\infty \sum _{h_1=0}^{\mathcal{E}_{II}-1}\sum _{q_0+q_1+\ldots+q_{\Xi _6}=h_1} \prod _{p_1} \binom{\mathcal{E}_{II}-1}{h_1} \binom{h_1}{ q_0,q_1,\ldots,q_{\Xi _6}} \Biggl(\frac{\Xi _6!}{p_1!\Xi _5^{\Xi _6-p_1+1}}\Biggl)^{q_{p_1}} \Biggl(\sum _{f_1=0}^{\infty}\Xi _4\Biggl)^{h_{1}} 
   \\
   & \times (-1)^{h_1} \mathcal{X} _6 \Biggl ( \frac{\Xi _3!\mathcal{X} _3!}{\Xi _2^{\Xi _3+1}} (\Xi _5+ \Xi _5 h_1)^{-(\mathcal{X} _3+1)} - \sum _{t_6=0}^{\Xi _3} \frac{\Xi _3! \mathcal{X} _7!}{t_6!\Xi _2^{\Xi _3-t_6+1}} (\Xi _2+\Xi _5 h_1+\Xi _5)^{-(\mathcal{X} _7+1)}\Biggl)\Biggl]
\end{align}
\vspace{0mm}
\end{figure*}

\subsection{Effective Secrecy Throughput Analysis}
Another secrecy performance metric is EST, which assures confidential average throughput measurements. EST is calculated by multiplying the target secrecy rate with the probability of successful data transmission that is defined as
\begin{align}
\label{eqn23}
\text{EST}=R_s(1-\text{SOP}).
\end{align}

\subsubsection{Scenario-I}
For colluding eavesdroppers, the formula of EST is expressed as
\begin{align}
\label{eqn43}
\text{EST}^{I}=R_s(1-\text{SOP}^{I}).
\end{align}
\subsubsection{Scenario-II}
For non-colluding eavesdroppers, the formula of EST is expressed as
\begin{align}
\label{eqn44}
\text{EST}^{II}=R_s(1-\text{SOP}^{II}).
\end{align}


\section{Numerical Results}
\begin{figure}[t!]
\vspace{0mm}
\centerline{\includegraphics[width=0.40\textwidth]{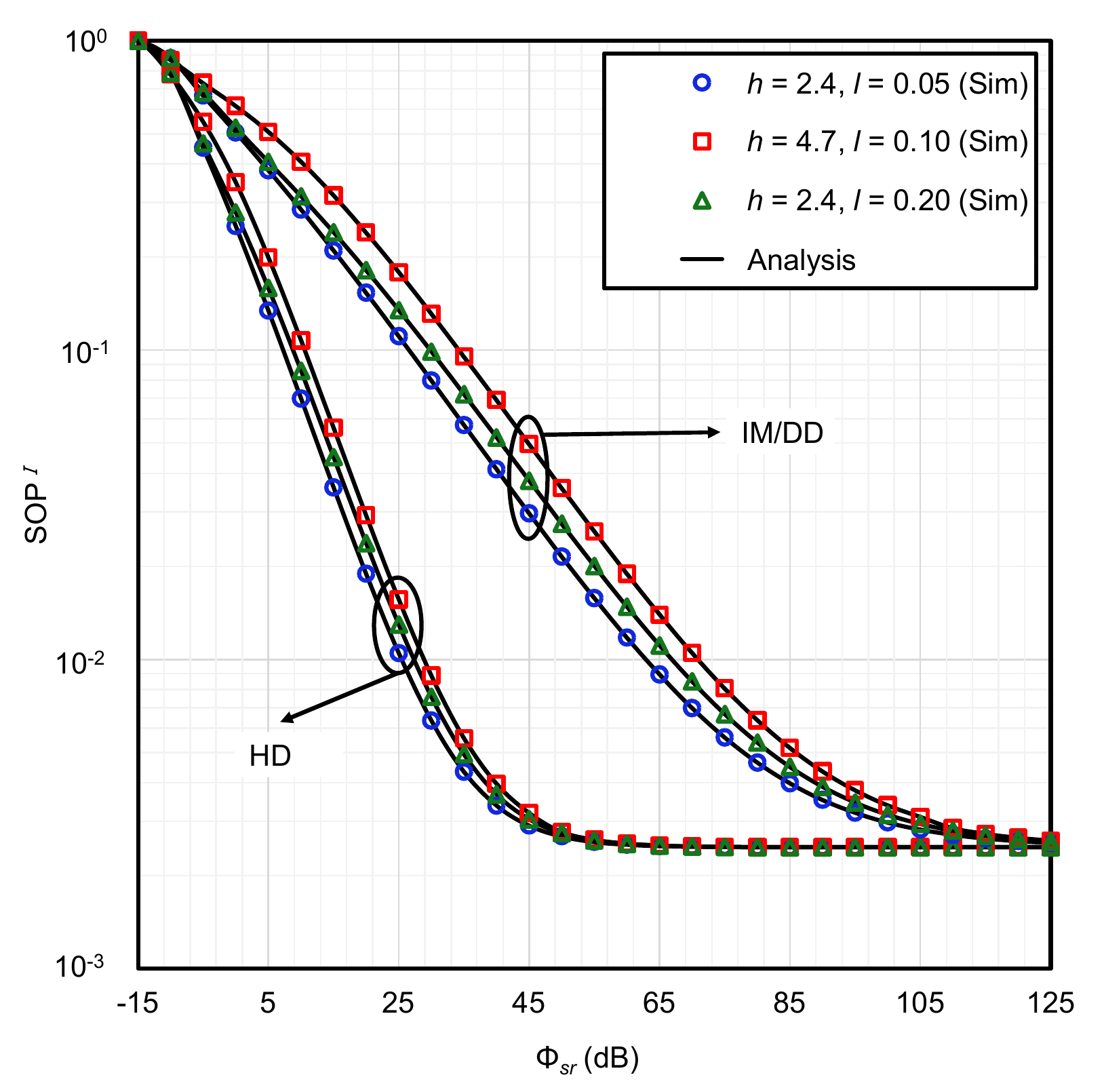}}
    \vspace{-3mm}
    \caption{The $\text{SOP}^{I}$ versus $\Phi _{sr}$ for selected values of $h$, $l$, and $\epsilon$ with $\xi=0.8$, $G_d=G_b=G_e=2$, $\kappa _d=\kappa _b=\kappa _e=1$, $\mu _d=\mu _b=\mu _e=1$, $m_d=m_b=m_e=2$, $\eta_{r} =0.7$, $\mathcal{E}_{I}=1$, $P_b=20$ dB, $\Phi _{rd}=15$ dB, $\Phi _{br}=1$, $\Phi _{re}=0$ dB, and $R_s=0.05$ bits/sec/Hz.}
    \label{Fig:1}
    \vspace{0mm}
\end{figure}

Selected simulation results due to mixed UOWC-RF model with the considered two different eavesdropping scenarios (colluding and non-colluding) are demonstrated and analyzed in this section, utilizing the obtained closed-form expressions of \eqref{eqn29}, \eqref{eqn52}, \eqref{eqn42}, \eqref{eqn58}, \eqref{eqn43}, and \eqref{eqn44}. In the simulation analysis, the impacts of detection techniques, various UWT scenarios, pointing error, fading and shadowing severity, number of diversity branches, number of eavesdroppers, energy conversion efficiency, power of power beacon, target secrecy rate, and average SNR values on secrecy performance are investigated. Similar to \cite{tania2022combined}, we assume that the RF links have the following parameters: $G_d=G_b=G_e$, $\kappa _d=\kappa _b=\kappa _e$, $\mu _d=\mu _b=\mu _e$, $m_d=m_b=m_e$, $\eta_{r} =0.7$, $P_b=20$ dB, and $R_s=0.05$ bits/sec/Hz, unless specified otherwise. On the other hand, the values of $h$ and $l$ utilized in the first hop are set according to \cite{zedini2019unified} and inherited from Table I and Table II. Note that Figs. \ref{Fig:1}-\ref{Fig:4} and Figs. \ref{Fig:6}-\ref{Fig:10} are illustrated for scenario-I (colluding eavesdropping mode) whereas Figs. \ref{Fig:5} and \ref{Fig:11} are depicted for scenario-II (non-colluding eavesdropping mode). Moreover, a fair comparison between colluding and non-colluding eavesdropping attacks is also presented in Figs. \ref{Fig:12} and \ref{Fig:13} in terms of EST and SOP analysis, respectively. The Monte-Carlo (MC) simulated results, with $10^{8}$ channel realizations and the analytical results, are utilized to generate all the graphs. Furthermore, the validity of our developed expressions is confirmed by the excellent match between theoretical and simulated results.


\subsection{Impact of UOWC Link Parameters}
To assess the severity level of different turbulence (i.e, air bubbles level and temperature gradients) conditions, $\text{SOP}^{I}$ is plotted as a function of $\Phi_{sr}$ in Fig. \ref{Fig:1}. It can be observed that secrecy performance decreases as the values of $h$ and $l$ increase. This is because an increase in the level of air bubbles and/or temperature gradient adversely affects the scintillation index, which has a negative effect on the SOP performance. Therefore, it can be observed that utilizing the HD technique instead of the IM/DD technique significantly improves secrecy performance. This is as expected since the HD technique, relative to the IM/DD technique, can more easily overcome the impacts of turbulence conditions as testified in \cite{9622195}. As noticed from the figure, the analytical results perfectly match with the MC simulation (denoted by markers). Additionally, it is demonstrated that SOP declines as $\Phi_{sr}$ increases proving that increasing $\Phi_{sr}$ enhances the secrecy performance.

\begin{figure}[!ht]
\vspace{-3.5mm}
\centerline{\includegraphics[width=0.40\textwidth]{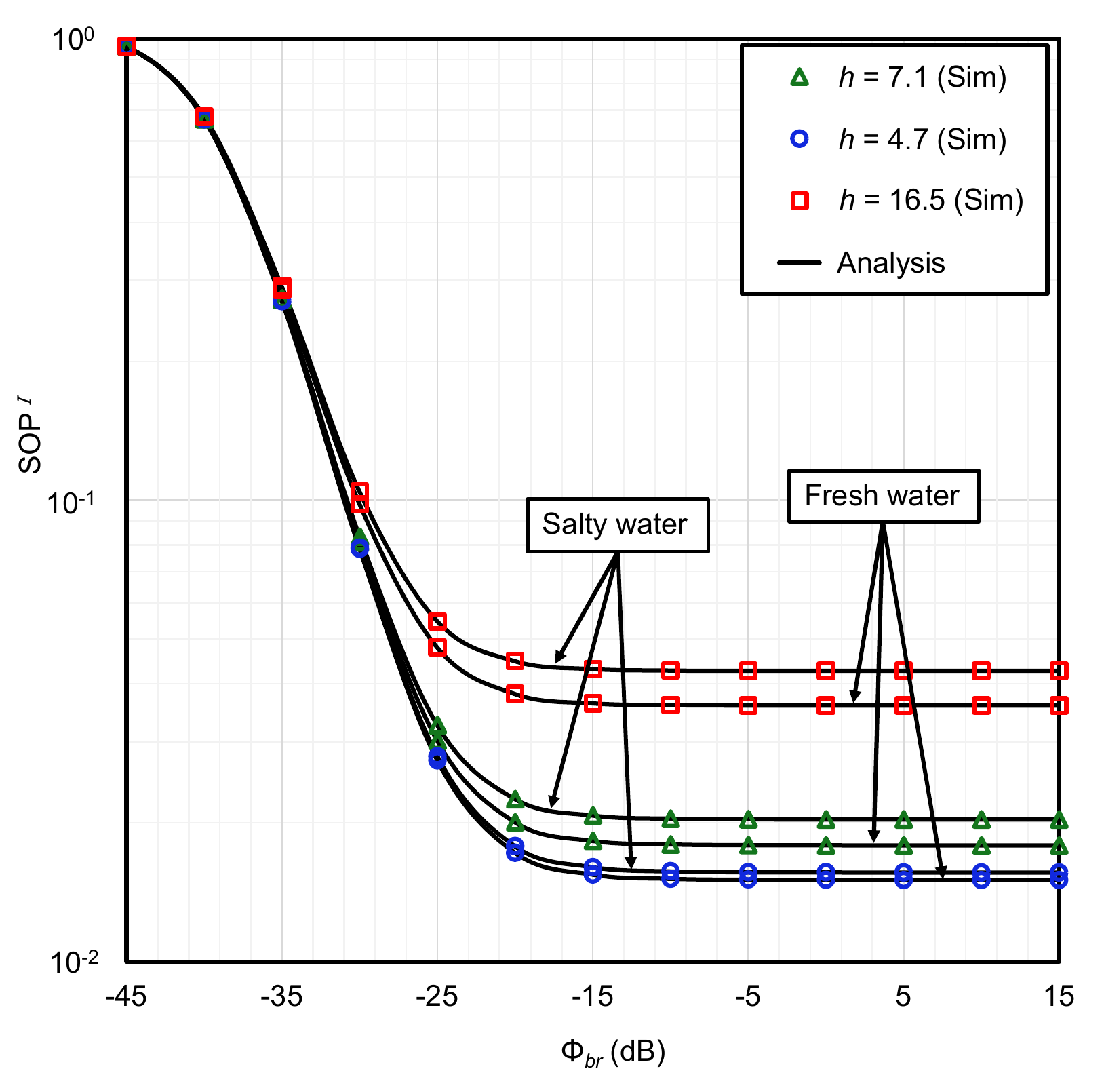}}
    \vspace{-3mm}
    \caption{The $\text{SOP}^{I}$ versus $\Phi _{br}$ for selected values of $h$ for both salty and fresh water with $\epsilon =1$, $\xi=0.8$, $G_d=G_b=G_e=2$, $\kappa _d=\kappa _b=\kappa _e=1$, $\mu _d=\mu _b=\mu _e=1$, $m_d=m_b=m_e=2$, $\eta_{r} =0.7$, $\mathcal{E}_{I}=1$, $P_b=20$ dB, $\Phi _{sr}=20$ dB, $\Phi _{rd}=10$ dB, $\Phi _{re}=-10$ dB, and $R_s=0.05$ bits/sec/Hz.}
    \label{Fig:2}
    \vspace{-3.5mm}
\end{figure}

In Fig. \ref{Fig:2}, $\text{SOP}^{I}$ performance is analyzed graphically concerning $\Phi_{br}$ under uniform temperature conditions and varying air bubbles levels due to the UOWC link.

It is possible to conclude that secrecy performance increases as the turbulence severity decreases. Water salinity also has an impact on secrecy performance though it is not as significant as the UWT scenarios. Increased value of $\Phi_{br}$ results in a considerable decline of SOP. This is because as $\Phi_{br}$ increases, the strength of $\mathcal{B-R}$ link improves, leading to a better SOP performance. However, a secrecy outage floor is observed in the figure after a certain value of $\Phi_{br}$. This is due to the fact that the secrecy performance is influenced by the worse hop in mixed UOWC-RF model implying the secrecy capacity is affected by the second hop in such a situation, which is not enhanced.

\begin{figure}[!t]
\vspace{0mm}
\centerline{\includegraphics[width=0.40\textwidth]{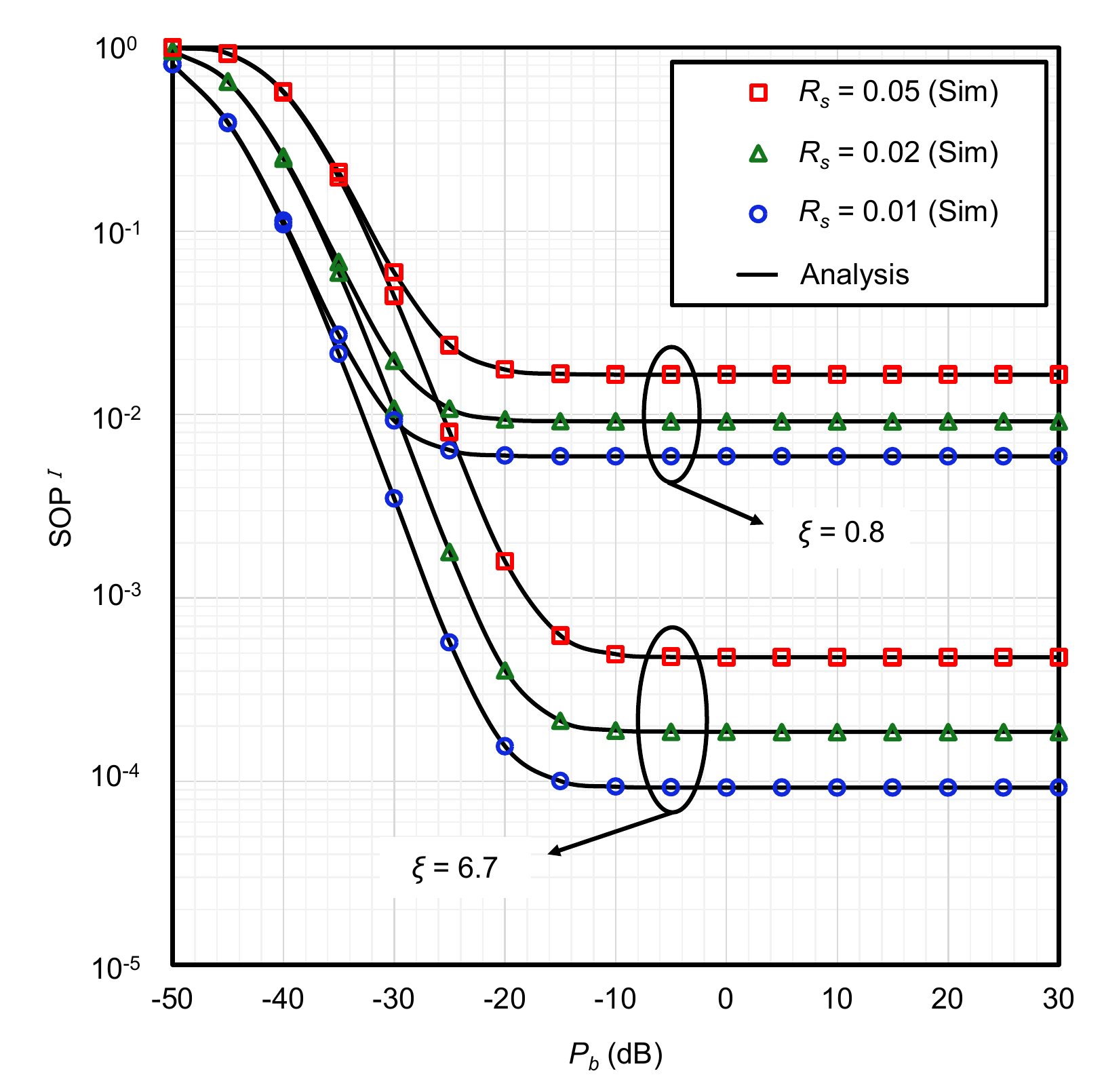}}
    \vspace{-3mm}
    \caption{The $\text{SOP}^{I}$ versus $P_b$ (dB) for selected values of $\xi$ and $R_s$ with $h=2.4$, $l=0.05$, $\epsilon =1$, $G_d=G_b=G_e=2$, $\kappa _d=\kappa _b=\kappa _e=1$, $\mu _d=\mu _b=\mu _e=1$, $m_d=m_b=m_e=2$, $\eta_{r} =0.7$, $\mathcal{E}_{I}=1$, $\Phi _{sr}=20$ dB, $\Phi _{rd}=30$ dB, $\Phi _{br}=1$ dB, and $\Phi _{re}=-10$ dB.}
    \label{Fig:3}
    \vspace{-3.5mm}
\end{figure}

Fig. \ref{Fig:3} depicts $\text{SOP}^{I}$ against $P_b$ under varying values of $R_{s}$ to address the influence of pointing errors in UOWC link. 
As expected, the higher value of $\xi$ (i.e., $\xi$=6.7) exhibits a better SOP performance than that of lower $\xi$ (i.e., $\xi$=0.8). This is due to the fact that smaller $\xi$ produces larger pointing errors in the receiver, which significantly distorts the signal that is received. As a result, the signal strength of $\mathcal{S-R}$ link deteriorates and the system's secrecy performance weakens. In a similar context, it is realized that the SOP value falls significantly with decreasing $R_s$. This is because $C_{sr}$ must be greater than $R_s$ to achieve secure networking over the UOWC-RF link. However, increasing $R_s$ increases the probability that $C_{sr}$ will fall below $R_s$ thereby raising the SOP value and as a result deteriorating the secrecy performance.
\subsection{Impact of RF Parameters}
\begin{figure}[t!]
\vspace{0mm}
\centerline{\includegraphics[width=0.40\textwidth]{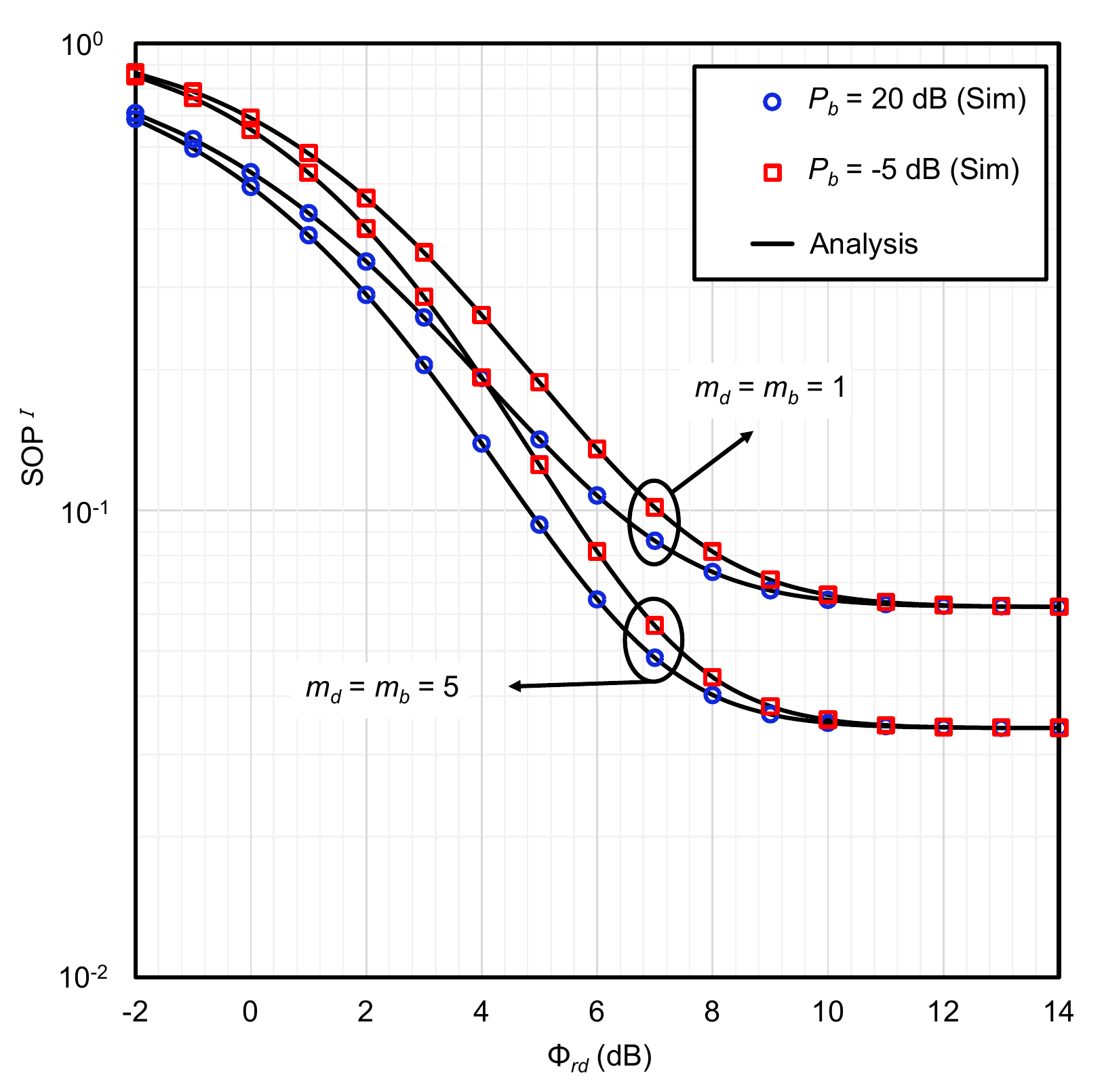}}
    \vspace{-3mm}
    \caption{The $\text{SOP}^{I}$ versus $\Phi _{rd}$ for selected values of $P_b$, $m_d$, and $m_b$ with $h=2.4$, $l=0.05$, $\epsilon=1$, $\xi =0.8$, $G_d=G_b=G_e=2$, $\kappa _d=\kappa _b=\kappa _e=1$, $\mu _d=\mu _b=3$, $\mu _e=1$, $m_e=1$, $\eta_{r} =0.7$, $\mathcal{E}_I=1$, $\Phi _{sr}=15$ dB, $\Phi _{br}=1$ dB, $\Phi _{re}=0$ dB, and $R_s=0.05$ bits/sec/Hz.}
    \label{Fig:4}
    \vspace{-3.5mm}
\end{figure}

The performance analysis of $\text{SOP}^{I}$ has been elucidated in Fig. \ref{Fig:4} to analyze the impacts of shadowing severity and beacon's power for the considered UOWC-RF mixed model.
It is noticed that $\text{SOP}^{I}$ increases as the value of $m_d$ and $m_b$ decreases. It is due to the fact that lowering the values of $m_d$ and $m_b$ reflects a stronger shadowing impact on the legitimate channel; hence the secrecy performance behaves inversely. Similarly, it is observed that SOP performance is greatly improved with increasing $P_b$ from $-5$ dB to $20$ dB. This increase enhances the possibility of improved energy harvesting at the relay with higher transmission capacity. However, as $\Phi_{rd}$ increases, the SOP value declines noticeably demonstrating that increasing the value of $\Phi_{rd}$ improves secrecy performance.

\begin{figure}[t!]
\vspace{0mm}
\centerline{\includegraphics[width=0.40\textwidth]{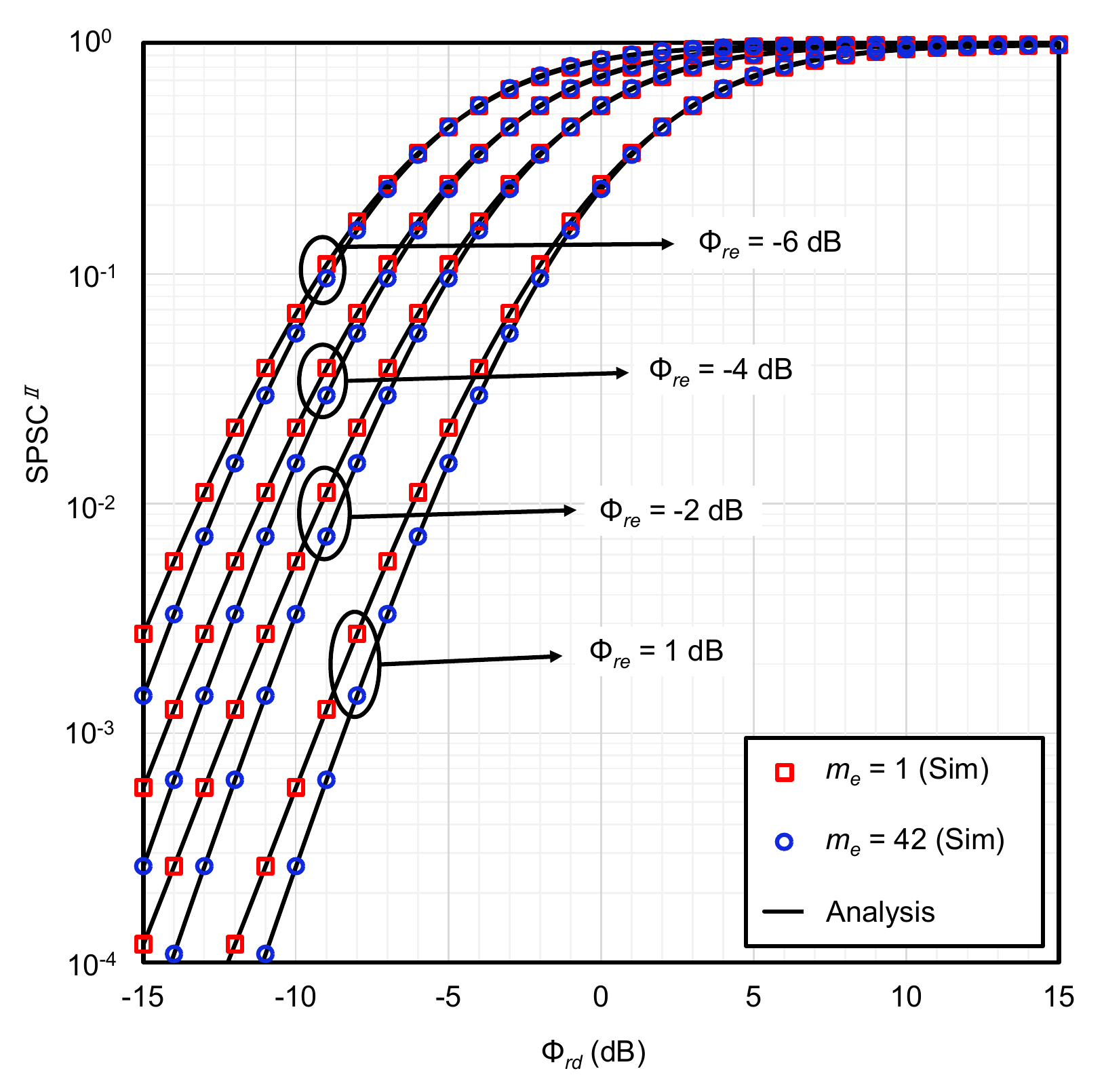}}
    \vspace{-3mm}
    \caption{The $\text{SPSC}^{II}$ versus $\Phi _{rd}$ for selected values of $\Phi _{re}$ and $m_e$ with $\mathcal{E}_{II}=2$, $G_d=G_e=2$, $\kappa _d=\kappa _e=1$, $\mu _d=\mu _e=1$, and $m_d=2$.}
    \label{Fig:5}
    \vspace{-3.5mm}
\end{figure}

In Fig. \ref{Fig:5}, $\text{SPSC}^{II}$ is plotted against $\Phi_{rd}$ demonstrating the effects of shadowing parameter of the eavesdropper channel.
\begin{figure}[b!]
\vspace{0mm}
\centerline{\includegraphics[width=0.40\textwidth]{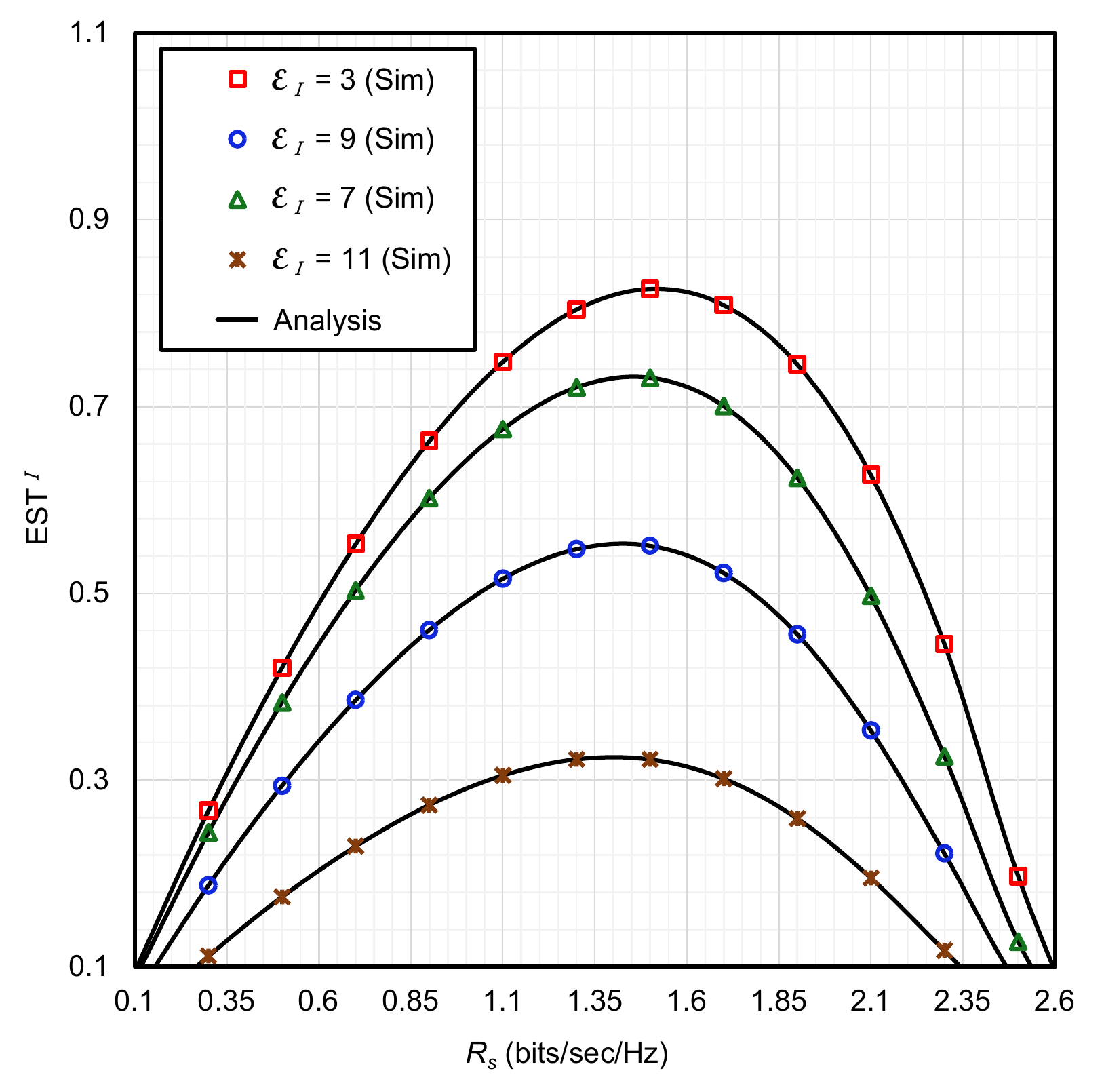}}
    \vspace{-3mm}
    \caption{The $\text{EST}^{I}$ versus $R_s$ for selected values of $\mathcal{E}_{I}$ with $h=2.4$, $l=0.05$, $\epsilon =1$, $\xi =0.8$, $G_d=G_b=G_e=2$, $\kappa _d=\kappa _b=\kappa _e=1$, $\mu _d=\mu _b=\mu _e=1$, $m_d=m_b=m_e=2$, $\eta_{r} =0.7$, $P_b=20$ dB, $\Phi _{sr}=15$ dB, $\Phi _{rd}=25$ dB, $\Phi _{br}=1$ dB, and $\Phi _{re}=0$ dB.}
    \label{Fig:6}
    \vspace{0mm}
\end{figure}
\begin{figure}[ht!]
\vspace{0mm}
\centerline{\includegraphics[width=0.40\textwidth]{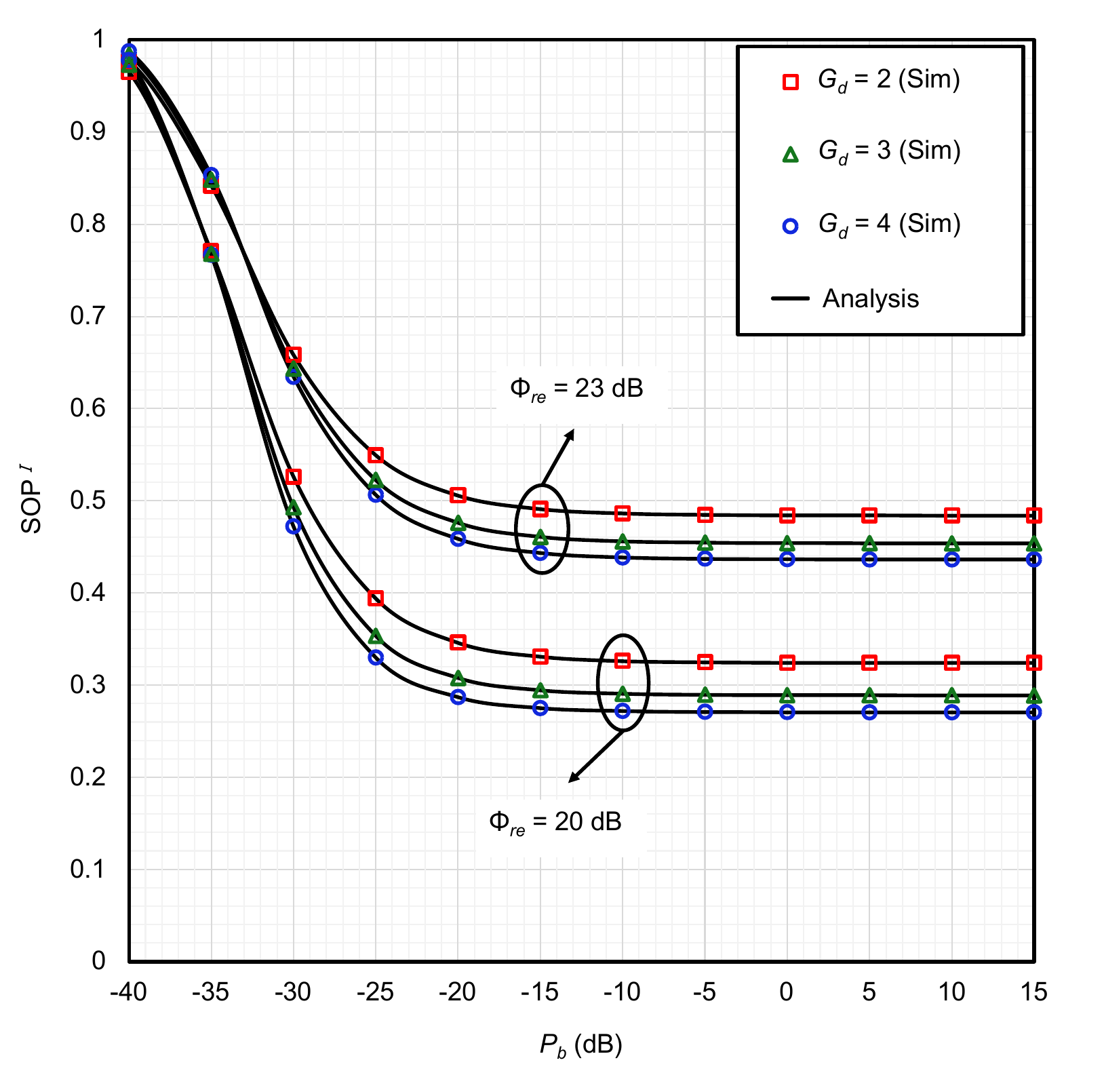}}
    \vspace{-3mm}
    \caption{The $\text{SOP}^{I}$ versus $P_b$ for selected values of $G_d$ and $\Phi _{re}$ with $h=2.4$, $l=0.05$, $\epsilon =2$, $\xi =0.8$,  $G_b=G_e=2$, $\kappa _d=\kappa _b=\kappa _e=1$, $\mu _d=\mu _b=\mu _e=1$, $m_d=m_b=m_e=2$, $\eta_{r} =0.7$, $\mathcal{E}_I=1$, $\Phi _{sr}=15$ dB, $\Phi _{rd}=25$ dB, $\Phi _{br}=1$ dB, and $R_s=0.05$ bits/sec/Hz.}
    \label{Fig:7}
    \vspace{-3.5mm}
\end{figure}
It can be observed that the lower the value of $m_{e}$, the better the SOP performance. This is because the SNR values of $\mathcal{R-E}$ link improve with the $m_{e}$ as the impact of shadowing is reduced while all other parameters remain constant. It is also realized that reducing the value of $\Phi_{re}$ provides better secrecy performance of the system. On the other hand, Fig. \ref{Fig:6} demonstrates the $\text{EST}^{I}$ of the UOWC-RF mixed system for selected values of $\mathcal{E}_I$ under various turbulence conditions to address the impact of colluding eavesdroppers and UWT severity.
Increasing the number of eavesdroppers decreases the secrecy performance, as noticed in the figure. It is expected since the probability of information leakage is enhanced drastically with the increase in eavesdroppers. Furthermore, the results of this figure reveal that EST improves with an increase in $R_s$ to a specific threshold ($R_{s}=1.55$ bits/sec/Hz) and subsequently declines with a further rise in $R_s$ as shown in TABLE I. This is because when $R_s$ is lower, it is possible to achieve the desired secrecy performance with fewer resources. However, as $R_s$ increases, greater resources are needed to counter the increased security risks, leading to decreased EST performance. In summary, the graph along with TABLE I illustrate stronger security measures offer diminishing returns in terms of EST performance.

\begin{table*}
\label{table1}
\centering
\caption{The $\text{EST}^{I}$ versus $R_s$ for selected values of $\mathcal{E}_{I}$}
\vspace{0mm}
\scalebox{.90}{
\begin{tabular}{|c|c|c|c|c|c|c|c|c|c|c|c|c|c|}
\hline
\begin{tabular}[c]{@{}c@{}}$R_s$\\ (bits/sec/Hz)\end{tabular}        & 1.1 & 1.15 & 1.2 & 1.25 & 1.3 & 1.35 & 1.4 & 1.45 & 1.5 & 1.55 & 1.6 & 1.65 & 1.7 \\ \hline 
\begin{tabular}[c]{@{}c@{}}$\mathcal{E}_{I}=3$\\ (Sim)\end{tabular}  & 0.74785                 & 0.76471                  & 0.77969                 & 0.79272                  & 0.80373                 & 0.81264                  & 0.81937                 & 0.82383                  & 0.82593                 & 0.82558                  & 0.82268                 & 0.81710                  & 0.80876                 \\ \hline 
\begin{tabular}[c]{@{}c@{}}$\mathcal{E}_{I}=7$\\ (Sim)\end{tabular}  & 0.67581                 & 0.68993                  & 0.70215                 & 0.71239                  & 0.72057                 & 0.72660                  & 0.73038                 & 0.73183                  & 0.73083                 & 0.72730                  & 0.72114                 & 0.71224                  & 0.70051                 \\ \hline 
\begin{tabular}[c]{@{}c@{}}$\mathcal{E}_{I}=9$\\ (Sim)\end{tabular}  & 0.51565                 & 0.52592                  & 0.53466                 & 0.54181                  & 0.54728                 & 0.55100                  & 0.55291                 & 0.55291                  & 0.55094                 & 0.54692                  & 0.54076                 & 0.53242                  & 0.52181                 \\ \hline 
\begin{tabular}[c]{@{}c@{}}$\mathcal{E}_{I}=11$\\ (Sim)\end{tabular} & 0.30500                 & 0.31078                  & 0.31561                 & 0.31945                  & 0.32224                 & 0.32394                  & 0.32451                 & 0.32389                  & 0.32204                 & 0.31891                  & 0.31447                 & 0.30868                  & 0.30151                              \\ \hline   
\end{tabular}}
\end{table*}

\begin{figure}[t!]
\vspace{0mm}
\centerline{\includegraphics[width=0.40\textwidth]{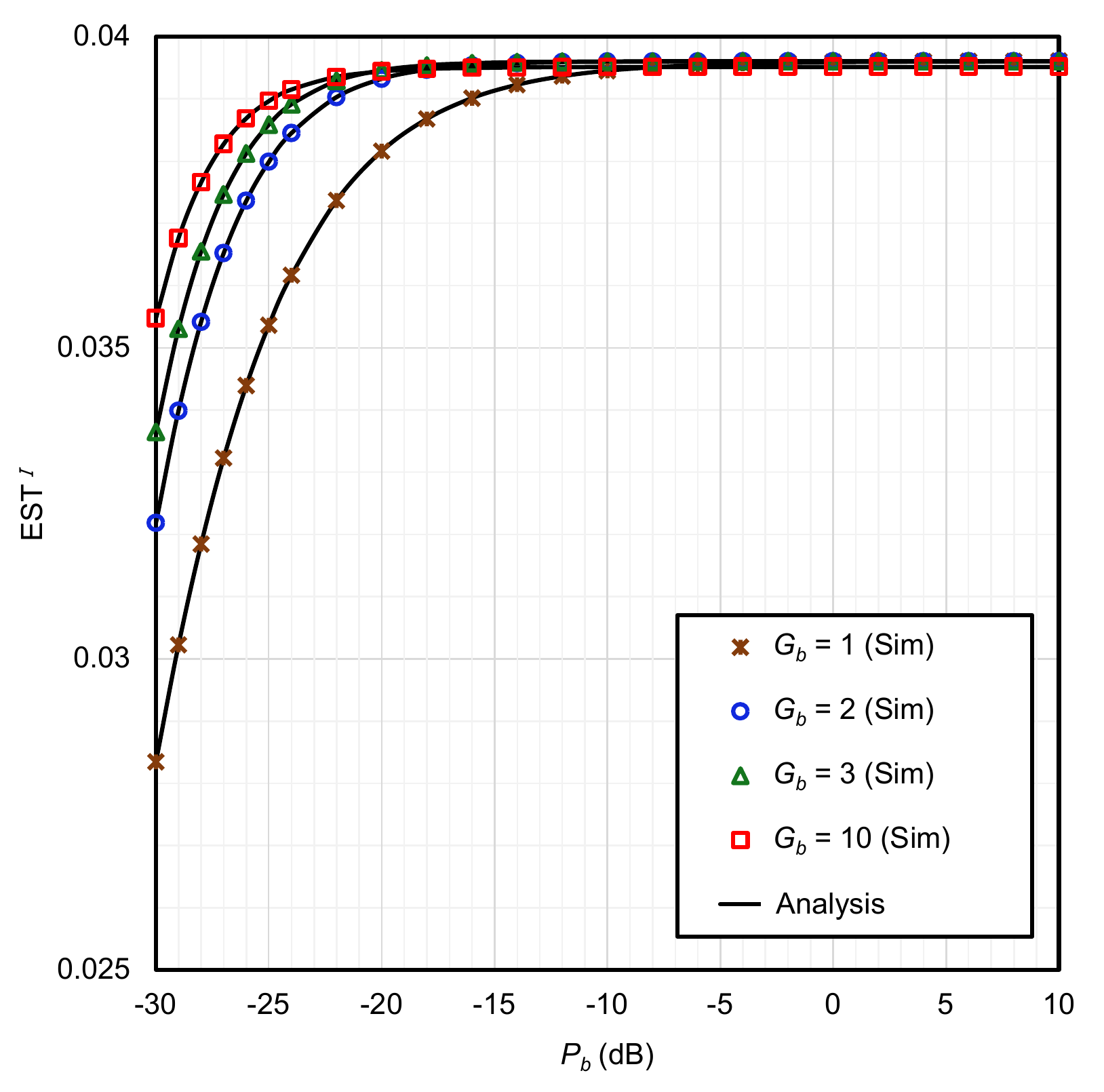}}
    \vspace{-3mm}
    \caption{The $\text{EST}^{I}$ versus $P_b$ for selected values of $G_b$ with $h=2.4$, $l=0.05$, $\epsilon=2$, $\xi =0.8$,  $G_d=G_e=2$, $\kappa _d=\kappa _b=\kappa _e=1$, $\mu _d=\mu _b=\mu _e=1$, $m_d=m_e=2$, $m_b=15$, $\eta_{r} =0.7$, $\mathcal{E}_I=2$, $\Phi _{sr}=15$ dB, $\Phi _{rd}=25$ dB, $\Phi _{br}=1$ dB, $\Phi _{re}=0$ dB, and $R_s=0.05$ bits/sec/Hz.}
    \label{Fig:8}
    \vspace{0mm}
\end{figure}
\begin{figure}[ht!]
\vspace{-3.5mm}
\centerline{\includegraphics[width=0.40\textwidth]{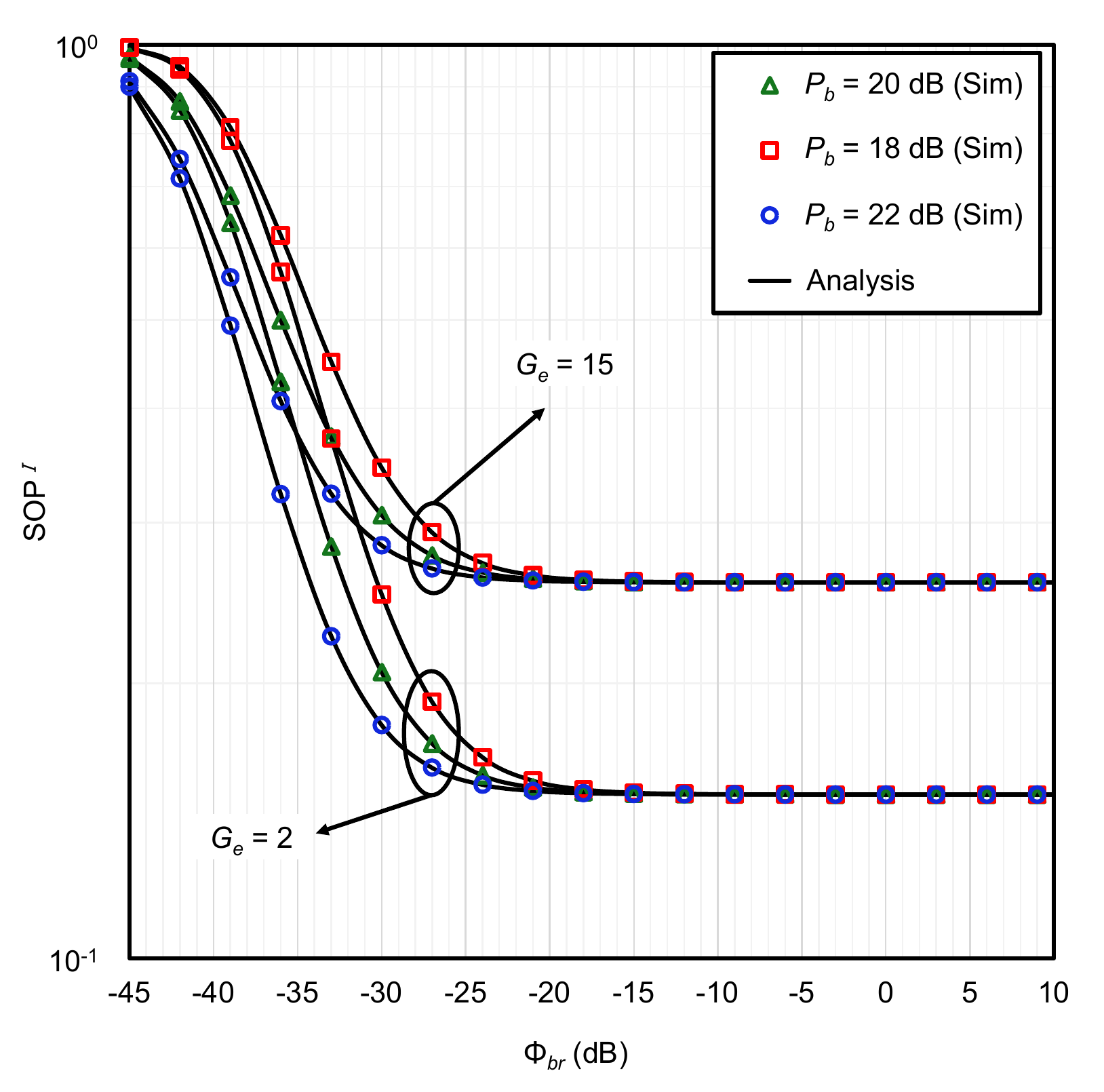}}
    \vspace{-3mm}
    \caption{The $\text{SOP}^{I}$ versus $\Phi _{br}$ for selected values of $P_b$ and $G_e$ with $h=2.4$, $l=0.05$, $\epsilon=2$, $\xi =0.8$, $G_d=G_b=2$, $\kappa _d=\kappa _b=\kappa _e=1$, $\mu _d=\mu _b=\mu _e=1$, $m_d=m_b=m_e=2$, $\eta_{r} =0.7$, $\mathcal{E}_I=2$, $\Phi _{sr}=20$ dB, $\Phi _{rd}=10$ dB, $\Phi _{re}=-10$ dB, and $R_s=0.05$ bits/sec/Hz.}
    \label{Fig:9}
    \vspace{-3.5mm}
\end{figure}

The influence of $G_{d}$ and $G_{b}$ on secrecy performance is investigated in Fig. \ref{Fig:7} and Fig. \ref{Fig:8}. Under both scenarios, it is observed that the proposed model exhibits better secrecy as the values of $G_{d}$ and $G_{b}$ increase. In other words, having more antennas at the destination and power-beacon is beneficial to strengthen the secrecy behavior. This is owing to the fact that implementing antenna diversity at the $\mathcal{D}$ and $\mathcal{B}$ ensures reliable communication, which leads to improved security. More importantly, an increase in $G_{b}$ increases the probability of exploiting more energy from the power-beacon to $\mathcal{R}$. On the other hand, it can be demonstrated that the higher value of $P_b$ ensures a good reception with a higher probability at $\mathcal{D}$. Note that when $P_b>-15$ dB, all such curves flatten out due to the dominance of system's first hop. The theoretical and simulation results of $\text{SOP}^{I}$ are compared in Fig. \ref{Fig:9} to analyze the impact of $G_{e}$ on secrecy performance. A clear decrease in the SOP performance is noticed with the increase of $G_{e}$. This is because when $G_{e}$ increases, more confidential information is susceptible to the eavesdroppers thereby increasing the chances of eavesdropping.
\begin{figure}[t!]
\vspace{0mm}
\centerline{\includegraphics[width=0.40\textwidth]{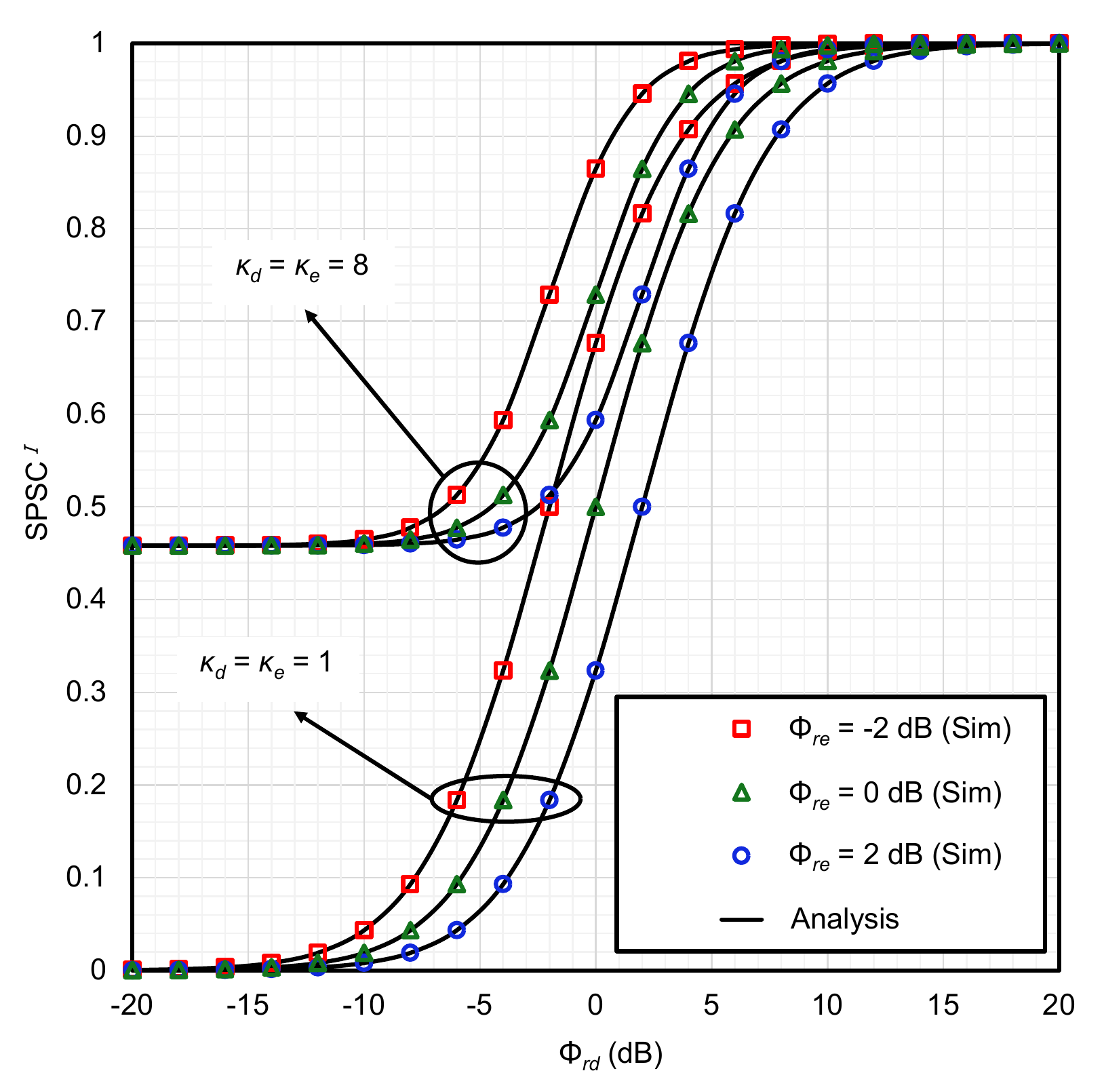}}
    \vspace{-3mm}
    \caption{The $\text{SPSC}^{I}$ versus $\Phi _{rd}$ for selected values of $\Phi _{re}$, $\kappa _d$, and $\kappa _e$ with $G_d=G_e=2$, $\mu _d=\mu _e=1$, $m_d=m_e=2$, and $\mathcal{E}_I=1$.}
    \label{Fig:10}
    \vspace{-3.5mm}
\end{figure}

\begin{figure}[t!]
\vspace{0mm}
\centerline{\includegraphics[width=0.40\textwidth]{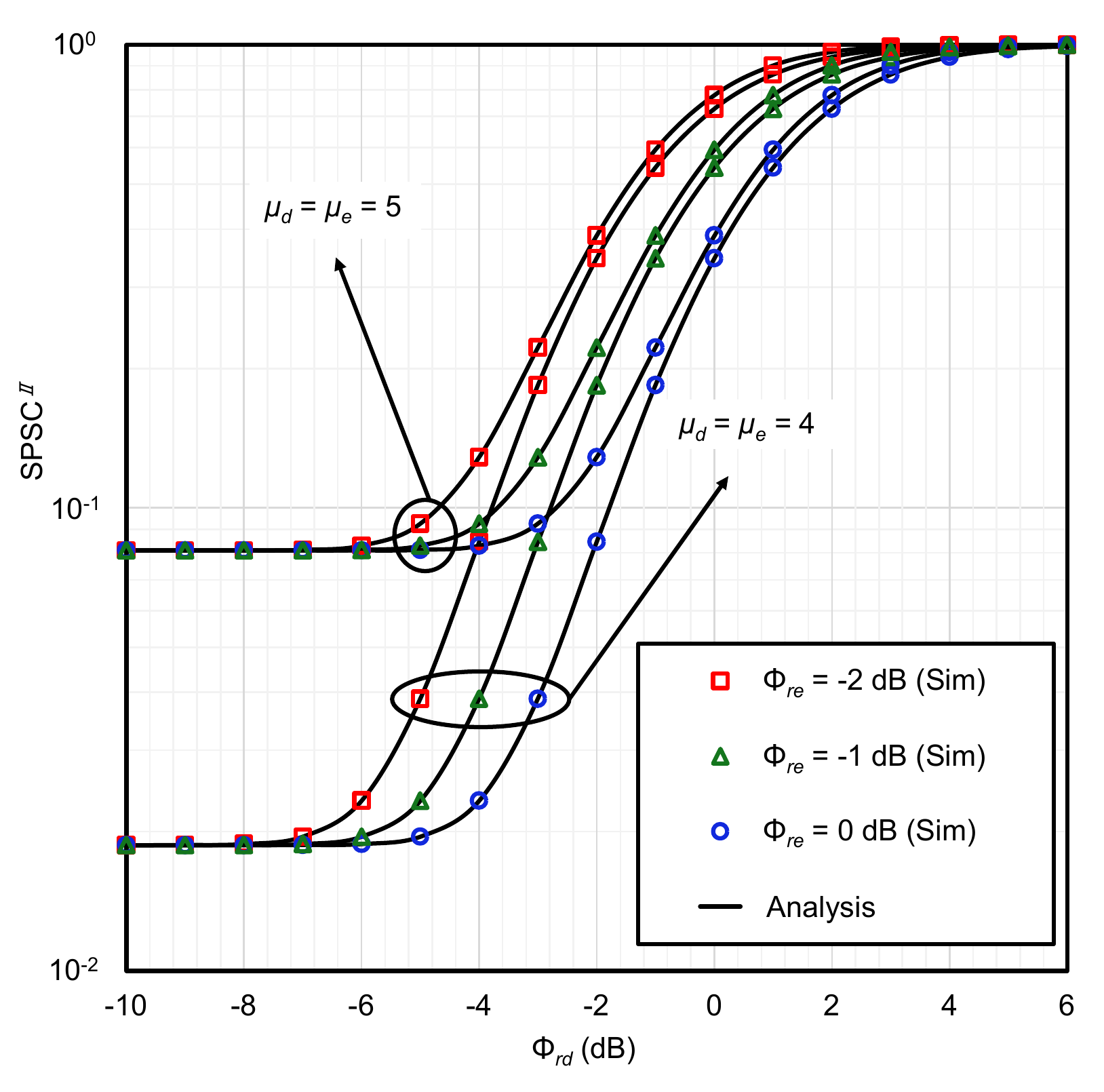}}
    \vspace{-3mm}
    \caption{The $\text{SPSC}^{II}$ $\Phi _{rd}$ for selected values of $\Phi _{re}$, $\mu _d$, and $\mu _e$ with $G_d=G_e=2$, $\kappa _d=\kappa _e=1$, $m_d=m_e=4$, and $\mathcal{E}_{II}=2$.}
    \label{Fig:11}
    \vspace{0mm}
\end{figure}

The effect of various fading conditions on secrecy performance is investigated in Figs. \ref{Fig:10}-\ref{Fig:11}, which are graphically represented in terms of the probability of SPSC analysis. The results in Fig. \ref{Fig:10} reveal that the probability of $\text{SPSC}^{I}$ greatly improves as the combination of $\kappa_{d}$ and $\kappa_{e}$ increases. This indicates that communication between $\mathcal{R}$ to $\mathcal{D}$ can be established more securely due to the increased fading severity. This is expected since the SNR of $\mathcal{R}-\mathcal{D}$ link is dependent on fading parameters as demonstrated in \eqref{eqn42}. Similar outcomes are also observed while comparing the results in Fig. \ref{Fig:11} since a good reception is obtained at $\mathcal{D}$ as the levels of $\mu_{d}$ and $\mu_{e}$ increase.
\subsection{Comparison of Colluding and Non-colluding Eavesdroppers Attack}

\begin{figure}[ht!]
\vspace{0mm}
\centerline{\includegraphics[width=0.40\textwidth]{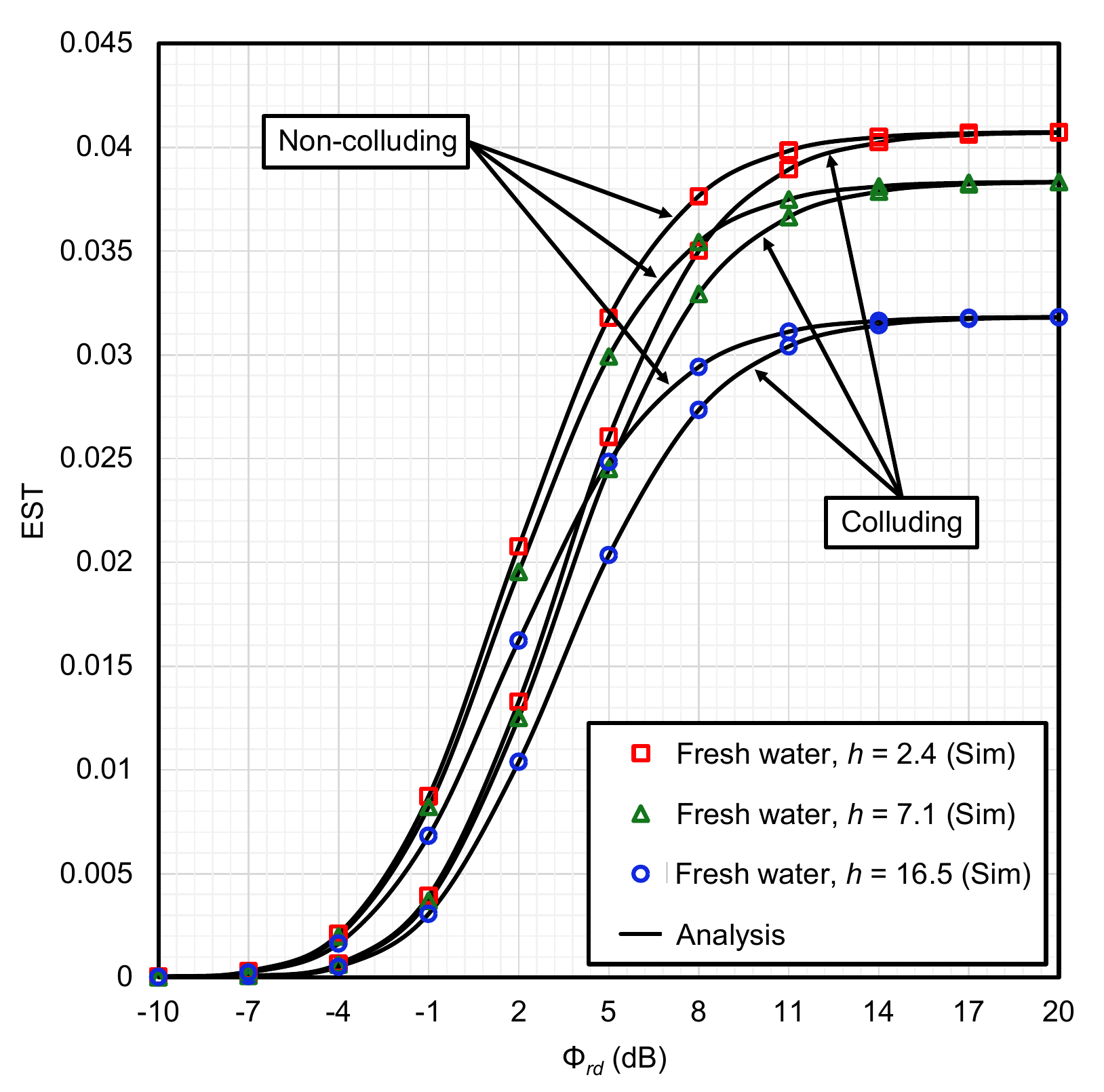}}
    \vspace{-3mm}
    \caption{The $\text{EST}$ versus $\Phi _{rd}$ for selected values of $h$ for fresh water with $\epsilon=2$, $\xi =0.8$, $G_d=G_b=G_e=2$, $\kappa _d=\kappa _b=\kappa _e=1$, $\mu _d=\mu _b=\mu _e=1$, $m_d=m_b=m_e=2$, $\eta_{r} =0.7$, $\mathcal{E}_I=\mathcal{E}_{II}=2$, $P_b=20$ dB, $\Phi _{sr}=15$ dB, $\Phi _{br}=1$ dB, $\Phi _{re}=0$ dB, and $R_s=0.05$ bits/sec/Hz.}
    \label{Fig:12}
    \vspace{-3.5mm}
\end{figure}

Fig. \ref{Fig:12} demonstrates the $\text{EST}$ analysis to compare the secrecy performance of colluding and non-colluding eavesdropping modes under numerous UWT conditions in thermally uniform freshwaters.
Since the non-colluding eavesdroppers obtain a greater EST value than the colluding mode, it can be inferred that the colluding eavesdropping scenario is more severe than that of the non-colluding scenario. In other words, due to the colluding scenario, eavesdropper's capabilities are improved, which leads to poorer secrecy performance. The reason for this is that while colluding eavesdroppers collaborate and corroborate their opinions to decode sensitive information, non-colluding eavesdroppers explore the confidential megssage independently. Furthermore, EST is greater in the presence of weaker UWT relative to stronger UWT conditions. The reason is the same as it was in Fig. \ref{Fig:2}.

\begin{figure}[t!]
\vspace{0mm}
\centerline{\includegraphics[width=0.40\textwidth]{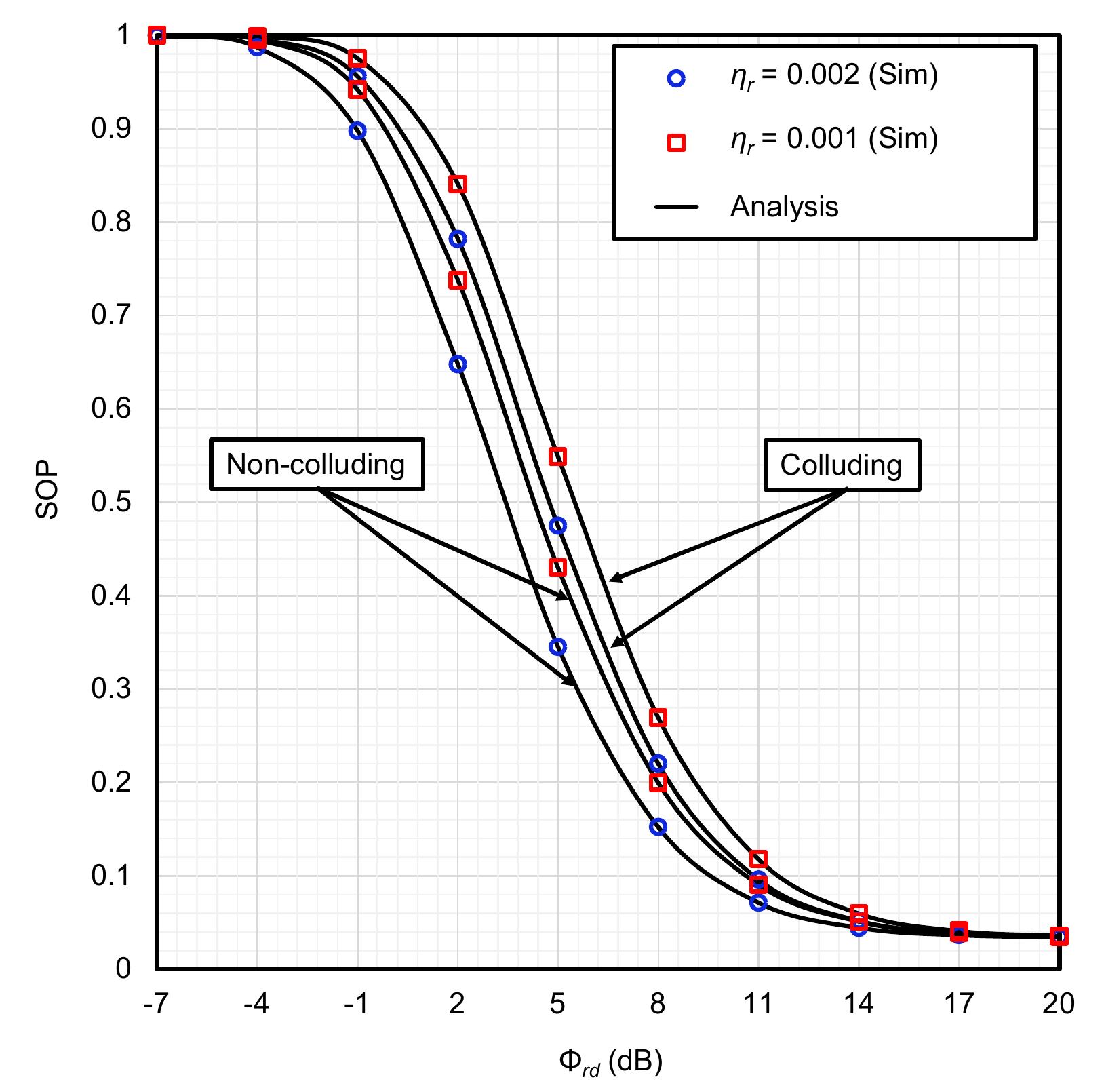}}
    \vspace{-3mm}
    \caption{The $\text{SOP}$ versus $\Phi _{rd}$ for selected values of $\eta_r$ with $h=2.4$, $l=0.05$, $\epsilon=1$, $\xi =0.8$, $G_d=G_b=G_e=2$, $\kappa _d=\kappa _b=\kappa _e=1$, $\mu _d=\mu _b=\mu _e=1$, $m_d=m_b=m_e=2$, $\mathcal{E}_I=\mathcal{E}_{II}=2$, $P_b=20$ dB, $\Phi _{sr}=15$ dB, $\Phi _{br}=1$, $\Phi _{re}=0$ dB, and $R_s=0.05$ bits/sec/Hz.}
    \label{Fig:13}
    \vspace{0mm}
\end{figure}

The {$\text{SOP}$} vs $\Phi _{rd}$ is illustrated in Fig. \ref{Fig:13}, which exhibits the impact of energy conversion efficiency $(\eta_{r})$.
It can be concluded that an increase in $\eta_{r}$ triggers a noticeable improvement in SOP performance. This is predictable since more harvested energy can be utilized due to the reliable information transfer in the second time slot. Moreover, it is observed that secrecy performance degrades significantly in the scenario of colluding eavesdroppers relative to the non-colluding eavesdroppers scenario.

\section{Conclusions}
This work analyses the secrecy performance of energy harvested relay-based UOWC-RF mixed network in the presence of two eavesdropping scenarios, i.e., colluding and non-colluding corresponding to \textit{Scenario-I} and \textit{Scenario-II}, respectively, and the key contribution is addressed initially via developing a mathematical model in terms of closed-form expressions of SOP, probability of SPSC and EST, and then validating the same via computer simulations. Numerical results reveal that whereas the secrecy performance is drastically influenced by UWT (based on air bubble level, temperature gradient, and water salinity), and pointing errors, the HD technique can guarantee a better secrecy throughput as opposed to the IM/DD technique. Besides, the secrecy performance is always dominated by the worse hop but it can be significantly enhanced via exploiting diversity at the power beacon and / or the destination along with increasing the power of the power beacon to facilitate more harvested energy at the relay. Finally, a comparative analysis between the two scenarios concludes that the attacks led by colluding eavesdroppers are more detrimental than that of non-colluding eavesdroppers, and preventing such collusion between the eavesdroppers must be given top-most priority by the design engineers while designing secure communication system networks.

\bibliographystyle{IEEEtran}
\bibliography{IEEEabrv,main.bib}

\end{document}